\documentclass[tighten,iop,apj]{emulateapj}

\usepackage{amsmath}
\usepackage{graphicx}
\usepackage{url}

\newcommand{\pd}{\partial}
\newcommand{\ud}{\ensuremath{\mathrm{d}}}

\begin{document}

\title{New Two-Dimensional Models of Supernova Explosions by the
  Neutrino-Heating Mechanism: Evidence for Different Instability
  Regimes in Collapsing Stellar Cores}

\shorttitle{Instability Regimes in Collapsing Stellar Cores}
\shortauthors{M\"uller et al.}

\author{Bernhard~M\"uller\altaffilmark{1}, Hans-Thomas~Janka\altaffilmark{1}, and Alexander~Heger\altaffilmark{2,3}}

\altaffiltext{1}{Max-Planck-Institut f\"ur Astrophysik,
  Karl-Schwarzschild-Str.~1, D-85748 Garching, Germany;
  bjmuellr@mpa-garching.mpg.de, thj@mpa-garching.mpg.de }

\altaffiltext{2}{School of Physics and Astronomy, University of Minnesota, 116 Church Street SE, Minneapolis, MN 55455, USA}

\altaffiltext{3}{Monash Center for Astrophysics, School of Mathematical Sciences, Building 28, Monash University, Victoria 3800, Australia, alexander.heger@monash.edu}

\begin{abstract}
The neutrino-driven explosion mechanism for core-collapse supernovae
in its modern flavor relies on the additional support of
hydrodynamical instabilities in achieving shock revival. Two possible
candidates, convection and the so-called standing accretion shock
instability (SASI), have been proposed for this role. In this paper,
we discuss new successful simulations of supernova explosions that
shed light on the relative importance of these two instabilities.
While convection has so far been observed
to grow first
in
self-consistent hydrodynamical models with multi-group neutrino
transport, we here present the first such simulation in which the SASI
grows faster while the development of convection is initially
inhibited. We illustrate the features of this SASI-dominated regime
using an explosion model of a $27 M_\odot$ progenitor, which is
contrasted with a convectively-dominated model of an $8.1 M_\odot$
progenitor with subsolar metallicity, whose early post-bounce behavior
is more in line with previous $11.2 M_\odot$ and $15 M_\odot$
explosion models.  We analyze the conditions discriminating between
the two different regimes, showing that a high mass-accretion rate and
a short advection time-scale are conducive for strong SASI activity.
We also briefly discuss some important factors for capturing the
SASI-driven regime, such as general relativity, the progenitor
structure, a nuclear equation of state leading to a compact
proto-neutron star, and the neutrino treatment.  Finally, we evaluate
possible implications of our findings for 2D and 3D supernova
simulations.
\end{abstract}


\section{Introduction}
\label{sec:intro}

Aspherical hydrodynamical instabilities play a paramount role in
core-collapse supernovae. In the the neutrino-driven explosion
mechanism in its modern guise, they are thought to be indispensable
agents for enhancing the efficiency of neutrino energy deposition in
the gain region sufficiently to eventually allow shock revival, but
their importance does not stop there. The asphericities that already
develop before the explosion is launched pre-determine the morphology
of the explosion in the later phases, and are thus a crucial factor
for the mixing instabilities in
supernovae \citep{arnett_89,kifonidis_03,kifonidis_06,hammer_10},
the development of pulsar kicks
\citep{janka_94,herant_95,burrows_96,scheck_04,scheck_06,wongwathanarat_10,nordhaus_10b,nordhaus_12},
and the remnant structure \citep{ggpk_10}.

In the purely hydrodynamical case (i.e.\ in the absence of magnetic
fields), two instabilities that influence the dynamics in the
supernova core in the pre-explosion phase have been identified.  
Already in the 1990s it was recognized that the gain layer can be unstable to
convection \citep{bethe_90} as the neutrino heating of the material
that is advected down to the cooling region close to the proto-neutron
star surface produces a negative entropy gradient. This was confirmed
by the first generation of multi-dimensional supernova models
\citep{herant_92,burrows_92,herant_94,burrows_95,janka_96,mueller_97}.
The latter indeed showed the development of violent convective overturn in
the heating region and suggested that the convective exchange of
strongly neutrino-heated material from close to the proto-neutron star with
cool post-shock matter in combination with the longer exposure of the accreted
material to neutrino heating \citep{buras_06b,murphy_08}
can provide a powerful means of
increasing the efficiency of neutrino heating, boosting the post-shock
pressure, and eventually reviving the stalled shock.

Another instability of a quite different nature was discovered by
\citet{blondin_03}, who observed that a standing accretion shock as
encountered in core-collapse supernovae may be unstable to large-scale
$\ell=1$ and $\ell=2$ oscillation modes even in the absence of neutrino
heating, a result that has been confirmed by further hydrodynamical
simulations in a number of follow-up studies in 2D and 3D
\citep{blondin_06,ohnishi_06,blondin_07,scheck_08,iwakami_08,iwakami_09,fernandez_09a,fernandez_10}.
Linear stability analyses provided a sound theoretical framework for
the understanding of this ``standing accretion shock instability''
and identified an amplification cycle of entropy/vorticity and acoustic
perturbations between the shock and the
proto-neutron star surface as the underlying mechanism for the
instability
(\citealp{foglizzo_02,foglizzo_06,foglizzo_07,yamasaki_07}).
While
\citet{blondin_06} and \citet{laming_07} have argued for the possibility of a
purely acoustic cycle, \citet{guilet_12} have recently given strong arguments
for an advective-acoustic cycle.
 Similar to convection, strong SASI activity
may improve the efficiency of neutrino heating by increasing the
average shock radius and therefore the residence time of matter in the
gain region \citep{scheck_08,marek_09}.

While convection and the SASI are well understood in the linear
regime, where they can be clearly differentiated, e.g.\ on the basis
of the dominant wavenumber, which is typically higher for the
convective modes, see \citep{foglizzo_06} and by means of their growth
behavior (monotonic vs. oscillatory).  The situation is far more
complicated in the non-linear regime where both instabilities can
interact with convection triggering secondary shock oscillations or
vice versa, and the presence of a strong $\ell=1$ or $\ell=2$ mode may
not necessarily be indicative of the SASI. The evidence for the SASI
as ``main'' instability from recent 2D explosion models using
elaborate multi-group transport
\citep{buras_06b,marek_09,bruenn_09,suwa_10,mueller_12} as well as
from the first tentative explorations in 3D \citep{takiwaki_12} is
therefore ambiguous: While they all show the presence of the strong
$\ell=1$ and $\ell=2$ shock oscillations characteristic of the SASI at
late times, convection appears to be
the ``primary'' instability in the sense that it always begins to grow first a few tens of
milliseconds after bounce. Convection also appears to be the
``primary'' instability in the 3D models of \citet{fryer_02} and
in recent 2D and 3D simulations based on different implementations of
a simple light-bulb approximation
\citep{murphy_08,fernandez_09b,nordhaus_10,hanke_11,burrows_12,murphy_12}.
The work of \citet{hanke_11} in particular suggests that for a certain
choice of parameters in such approximate descriptions may little
activity of low-$\ell$ modes in 3D, while the inverse turbulent
  energy cascade may still help to excite low-$\ell$ modes in 2D even
  though convection initially grows faster than the SASI. Except for
some gray transport models with fast (prescribed) proto-neutron star
contraction of \citet{scheck_08}, the primary growth of convection
seems to be a generic property of simulations that include neutrino
heating in some simplified or elaborate fashion. The apparent weight
of evidence has even led to the suggestion ``that the SASI is at most
a minor feature of supernova dynamics'' \citep{burrows_12}
and that neutrino-driven convection dominates in core-collapse
supernovae \citep{murphy_12}.

However, we shall demonstrate in this paper that such a conclusion is
more than premature, and that under favorable conditions the SASI
plays a major role for the dynamics in the supernova core. We present
new simulations of (incipient) explosions for two different progenitor
stars obtained with the general relativistic multi-group neutrino
hydrodynamics code \textsc{Vertex-CoCoNuT}. Based on these, we show
the existence of a ``SASI-dominated'' regime distinct from a
``convection-dominated'' regime for the growth of hydrodynamical
instabilities as a separate route towards the explosion. After briefly
introducing the numerical methods and the progenitors in
Section~\ref{sec:setup}, we discuss the morphology of the
multi-dimensional flow prior and during the incipient explosion and
quantitatively analyze the growth of the SASI and convection in the
different regimes in Section~\ref{sec:results}.  The implications of
our findings are discussed in Section~\ref{sec:conclusions}.

\begin{figure}
\plotone{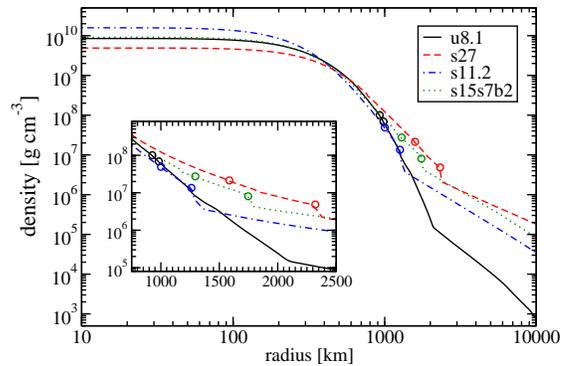}
\caption{Density profiles for model u8.1 (black solid
  line) and model s27.0 (red, dashed) covering the inner $10,000
  \ \mathrm{km}$. For comparison, profiles for the $11.2 M_\odot$
  progenitor s11.2 (blue, dash-dotted) of \citet{woosley_02} and the
  $15 M_\odot$ model s15s7b2 (green, dotted) of \citet{woosley_95} are also
  shown. Colored circles denote the edge of the iron core and the
  boundary between the silicon shell and the oxygen-enriched silicon
  shell.}
\label{fig:profiles}
\end{figure}

\section{Numerical Methods and Model Setup}
\label{sec:setup}
We perform axisymmetric (2D) core-collapse supernova simulations with
the general relativistic neutrino hydrodynamics code
\textsc{Vertex-CoCoNuT} \citep{mueller_10}. The hydrodynamics solver
\textsc{CoCoNuT} is a time-explicit, directionally unsplit
implementation of the piecewise parabolic method (PPM) for spherical
polar coordinates using a hybrid HLLC/HLLE Riemann solver, and relies
on the extended conformal flatness approximation (xCFC,
\citealp{cordero_09}) for the metric equations. \textsc{CoCoNuT} is
coupled to the neutrino transport module \textsc{Vertex}
\citep{rampp_02}, which employs a variable Eddington factor technique to
solve the moment equations for the neutrino energy and momentum
density with a closure provided by the formal solution of a simplified
Boltzmann equation.  In 2D, we resort to the ``ray-by-ray-plus''
approach \citep{buras_06_a,bruenn_06}, which assumes that the neutrino
distribution function is axially symmetric around the radial direction
(and hence implies a radial flux vector). \textsc{Vertex} includes the
velocity- and metric-dependent terms as well as non-isoenergetic
scattering processes, thus capturing the full complexity of
multi-group transport. For the interactions between neutrinos and the
matter we use an up-to-date set of opacities (see
\citealt{mueller_12}). For details about the implementation and a
discussion of the accuracy of our approach, we refer the reader to
\citet{mueller_10} and \cite{mueller_12}.

In this study, we consider two progenitors, namely the $27 M_\odot$
star model s27.0 of \citet{woosley_02} with solar metallicity, and an
$8.1 M_\odot$ model u8.1. Th latter is just above the critical mass
for the formation of an iron core and has an initial metallicity of
approximately $Z=10^{-4}$ \citep{HWZJ12}.\footnote{A model of $8.0
  M_\odot$ and $Z=10^{-4}$ did make an AGB star instead.}  The
structure of model u8.1 should be typical of such low-mass supernova
progenitors, and is dramatically different from stars even a few
tenths of a solar mass more massive.  The density profile more closely
resembles an asymptotic giant branch (AGB) star, or an
electron-capture supernova progenitor, than a typical massive star
structure in that it has a low-density ``envelope'' ($\rho < 1
\ \mathrm{g}\ \mathrm{cm}^{-3}$) directly on top of a dense core
($\rho > 10^6 \ \mathrm{g} \ \mathrm{cm}^{-3}$) of $1.38 M_\odot$,
with a transition region of only $0.03 M_\odot$, mostly the carbon
layer, in between. Outside the $1.26 M_\odot$ iron core are layers of
silicon (out to $1.30 M_\odot$), oxygen (to $1.36 M_\odot$), neon, and
carbon, with implosive energy generation due to silicon, oxygen, and
neon burning as high as $10^{17} \ \mathrm{erg} \ \mathrm{g}^{-1}
\ \mathrm{s}^{-1}$. Note that the structure of model u8.1 is specific
to such low-mass supernova progenitors, and not owing to the initial
metallicity of the model; a different initial metallicity would only
change the location and extent of the mass range between the AGB
channel and the ``normal'' channel of iron core evolution familiar
from more massive stars.

Model s27.0, by contrast, has a more massive and less compact iron
core of $1.5 M_\odot$ embedded in a thick silicon shell that reaches
out to $1.68 M_\odot$, where the transition to the oxygen-enriched
silicon shell is located. Compared to model u8.1, the density drops
far less rapidly outside the iron core. In order to
  better illustrate the different density structure of the two models,
  we show density profiles of the progenitors in
  Figure~\ref{fig:profiles}.

We use a numerical grid of $n_r \times n_\theta = 400 \times 128$
zones with non-equidistant radial spacing for both progenitors. Model
s27.0 was simulated using the equation of state (EoS) of
\citet{lattimer_91} with a value for the bulk incompressibility
modulus of nuclear matter of $K=220 \ \mathrm{MeV}$ (LS220), while the
softer variant with $K=180 \ \mathrm{MeV}$ (LS180) was chosen for
model u8.1. For a discussion of the validity of the latter EoS for
small-mass (baryonic mass $\lesssim 1.5 M_\odot$) proto-neutron stars
despite its marginal inconsistency with the $1.97 M_\odot$ neutron
star found by \citet{demorest_10}, see \citet{mueller_12}.\
Specifically, the mass-radius relation
for hot and cold neutron stars is very similar for neutron stars well below
the mass limit. As a consquence, both equations of state yield hardly any difference 
during the accretion phase \citep{myra_94,thompson_03}.

\begin{figure}
\plotone{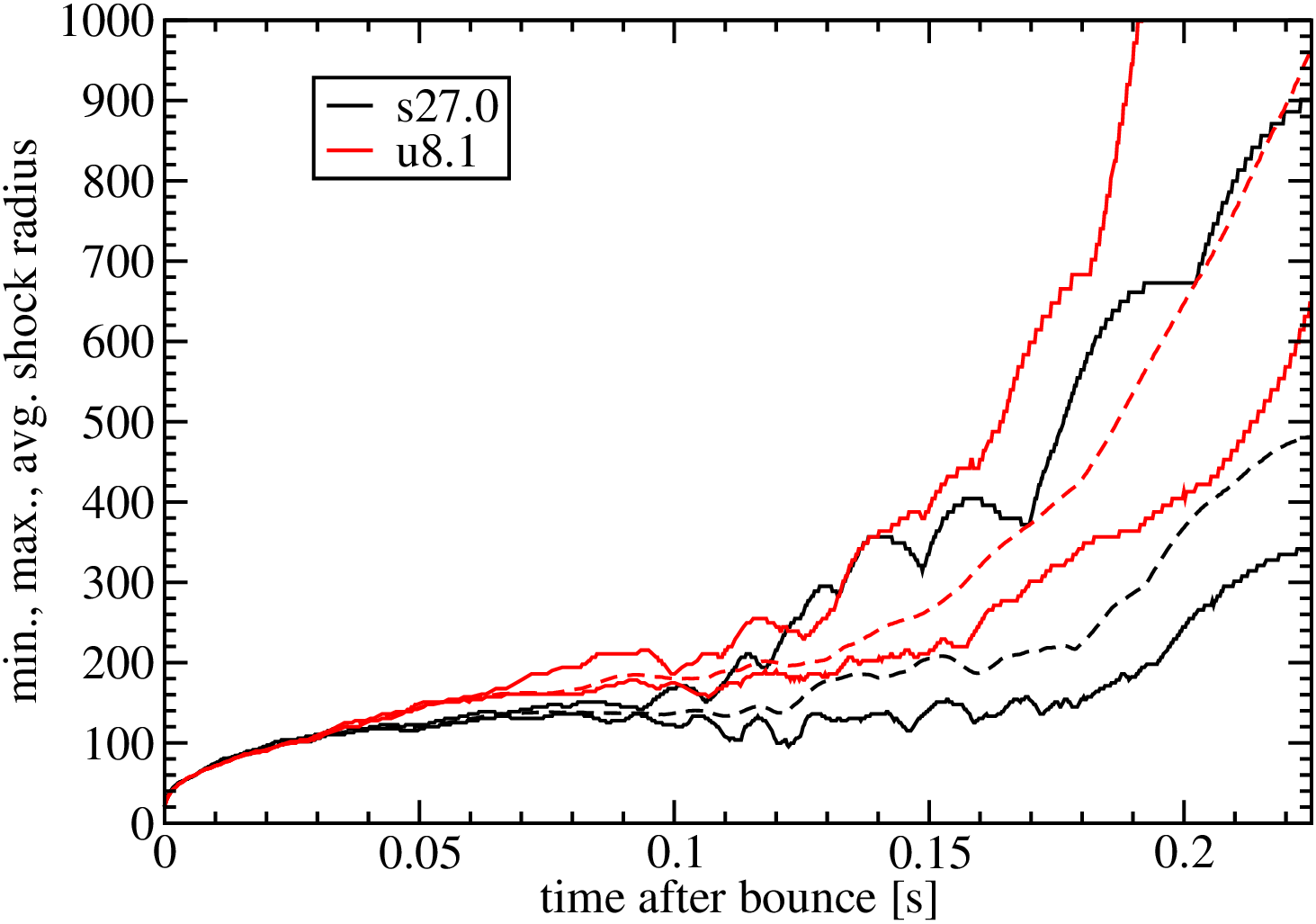}
\caption{Maximum, minimum (solid lines) and average (dashed lines)
  shock radius for model s27.0 (black lines) and u8.1 (red).}
\label{fig:shock}
\end{figure}

\begin{figure}
\plotone{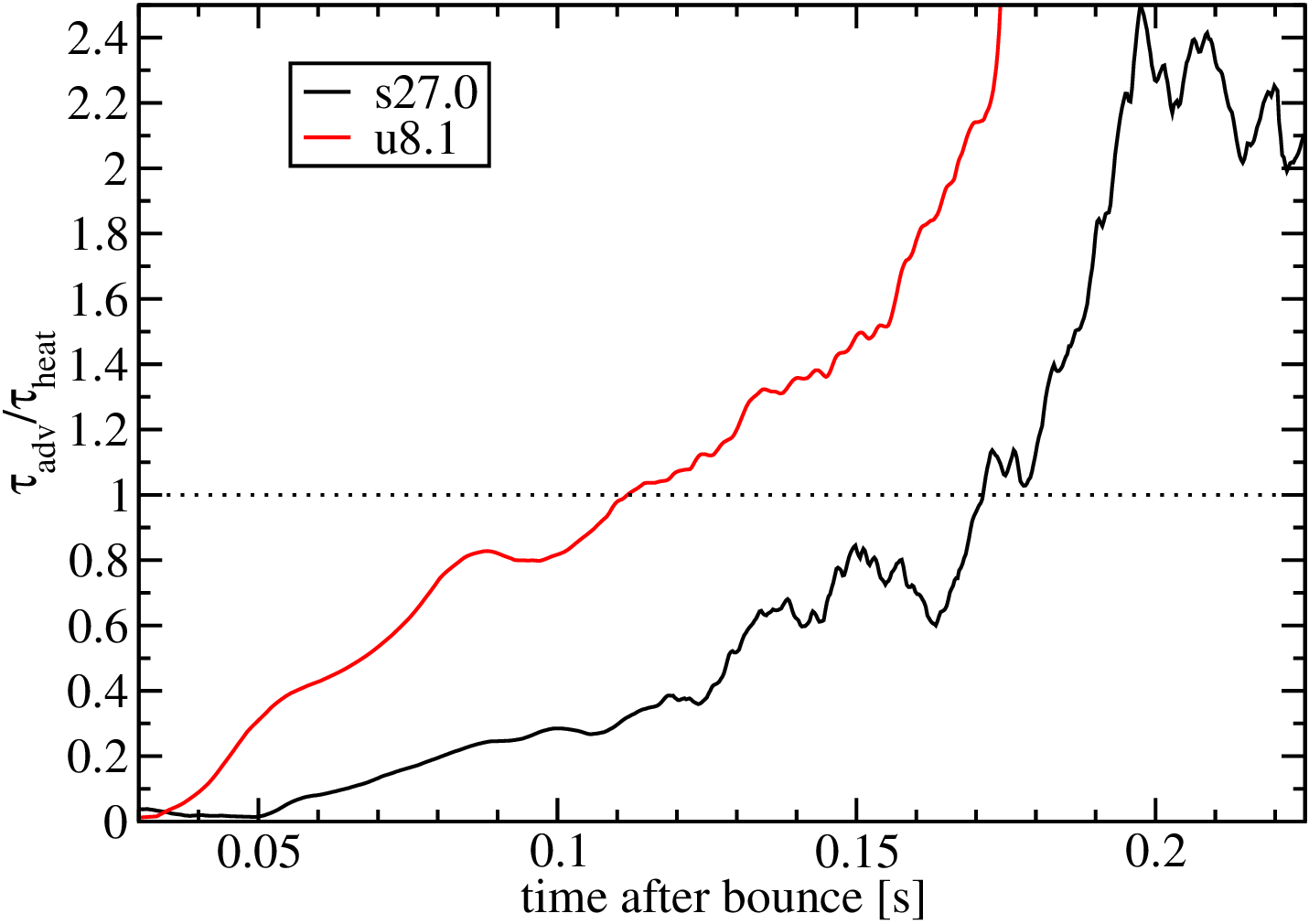}
\caption{The runaway criterion $\tau_\mathrm{adv}/\tau_\mathrm{heat}$
for model s27.0 (black) and model u8.1 (red). Both
$\tau_\mathrm{adv}$ and $\tau_\mathrm{heat}$ are evaluated as
in \citet{mueller_12}. Note that the curves have been smoothed
using a running average over $5 \ \mathrm{ms}$.
\label{fig:timescale_ratio}
}
\end{figure}

\section{Results}
\label{sec:results}

\begin{figure*}
\plottwo{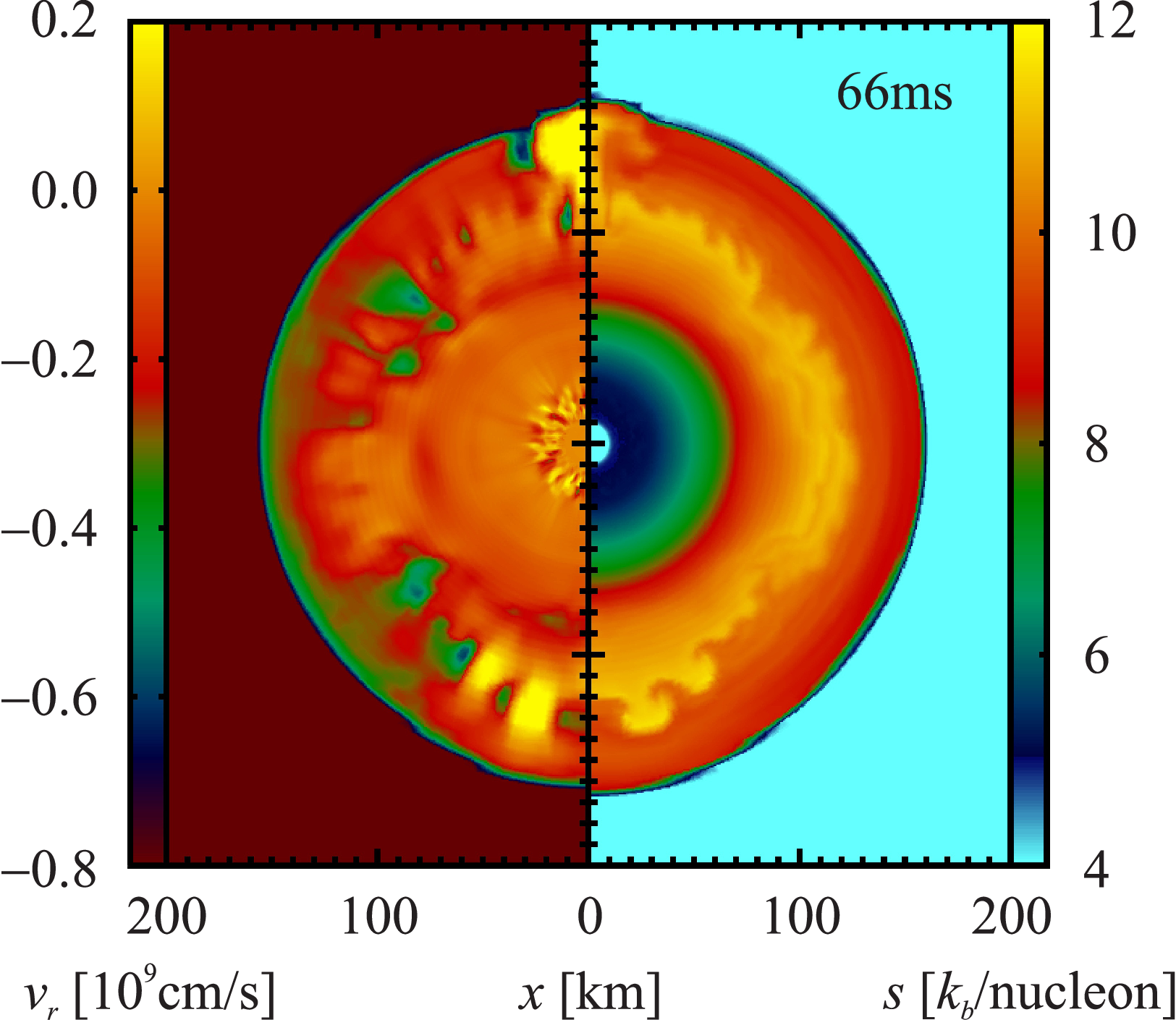}{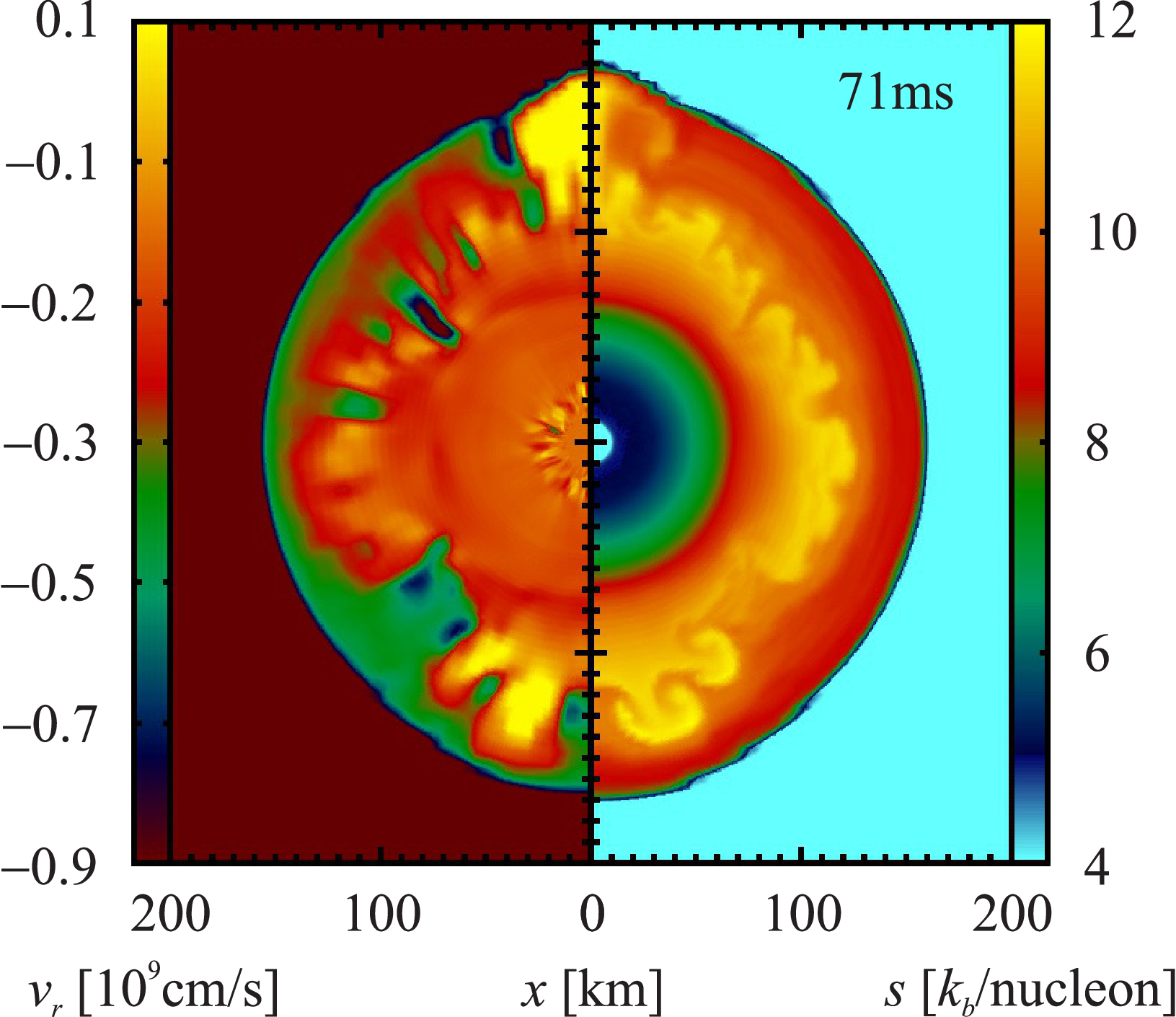}
\vspace{0.2in}
\plottwo{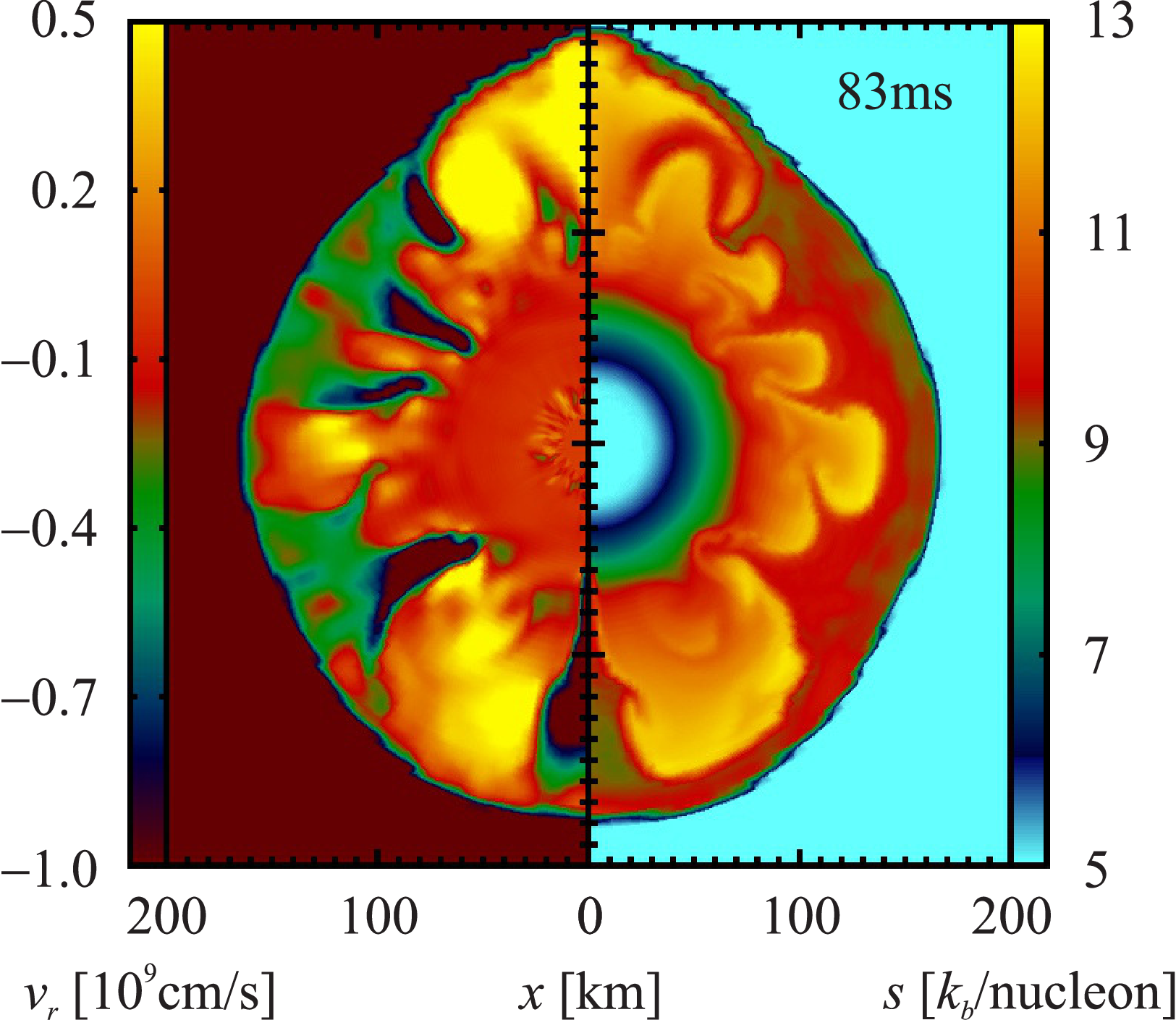}{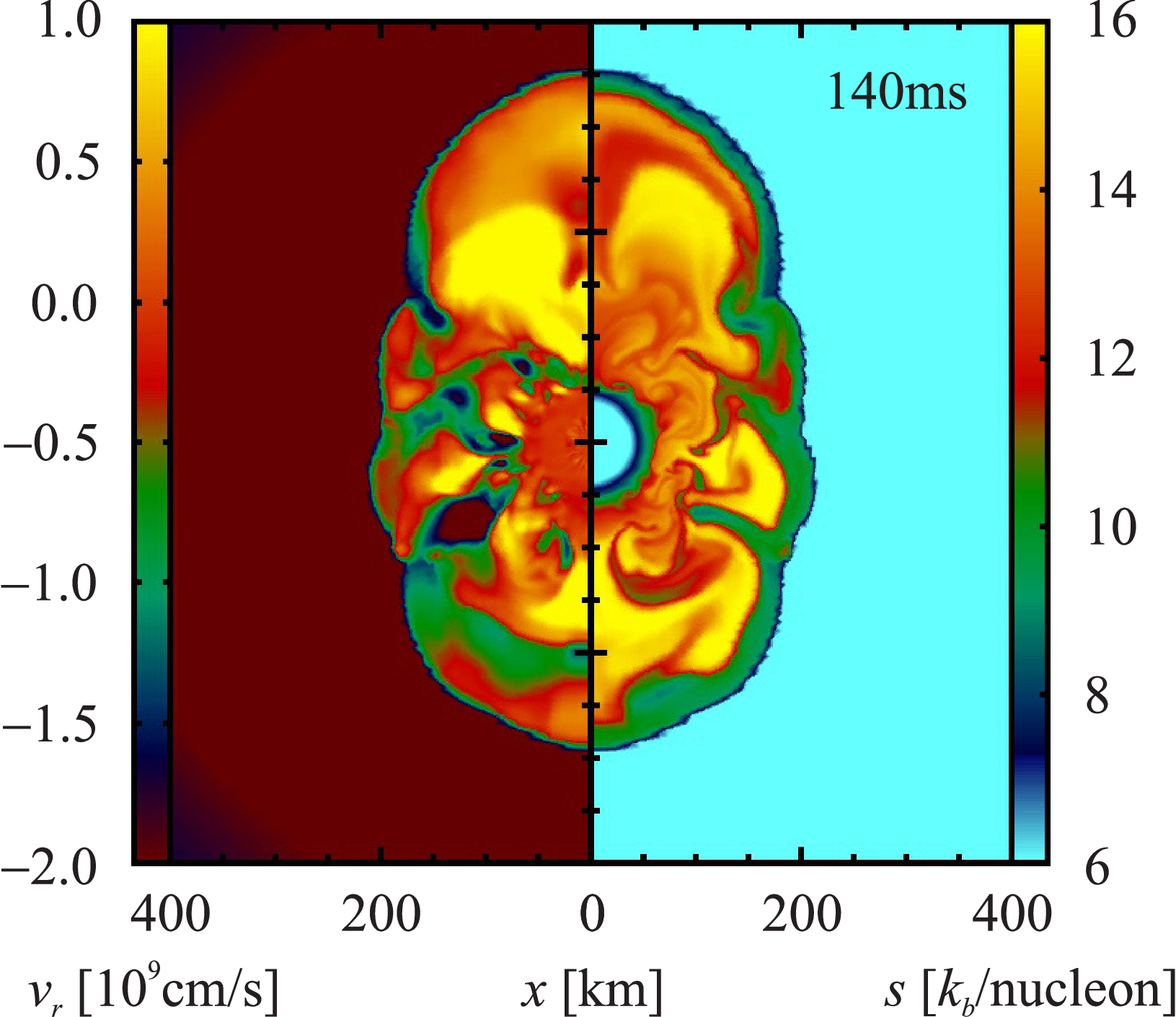}
\vspace{0.2in}
\plottwo{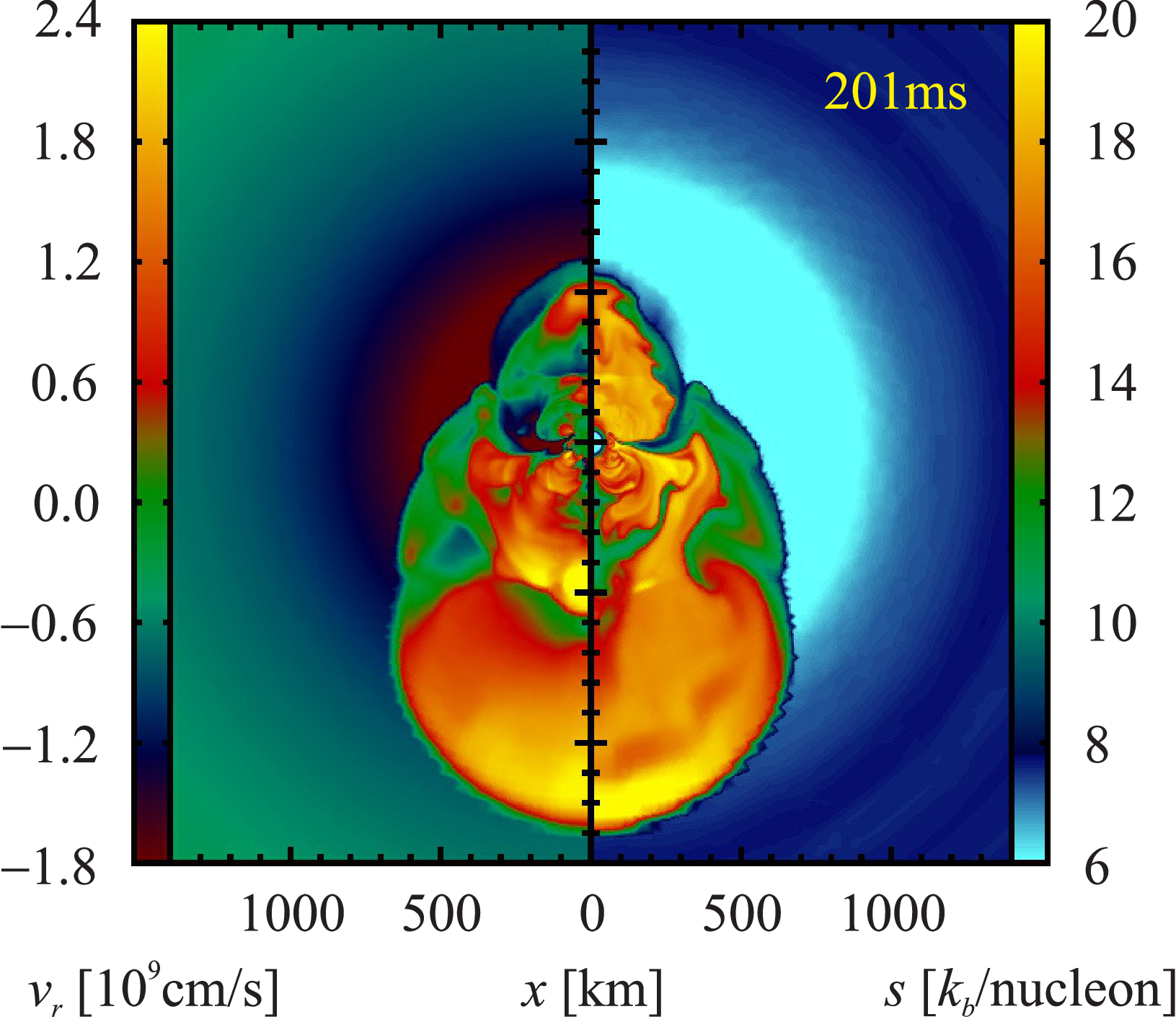}{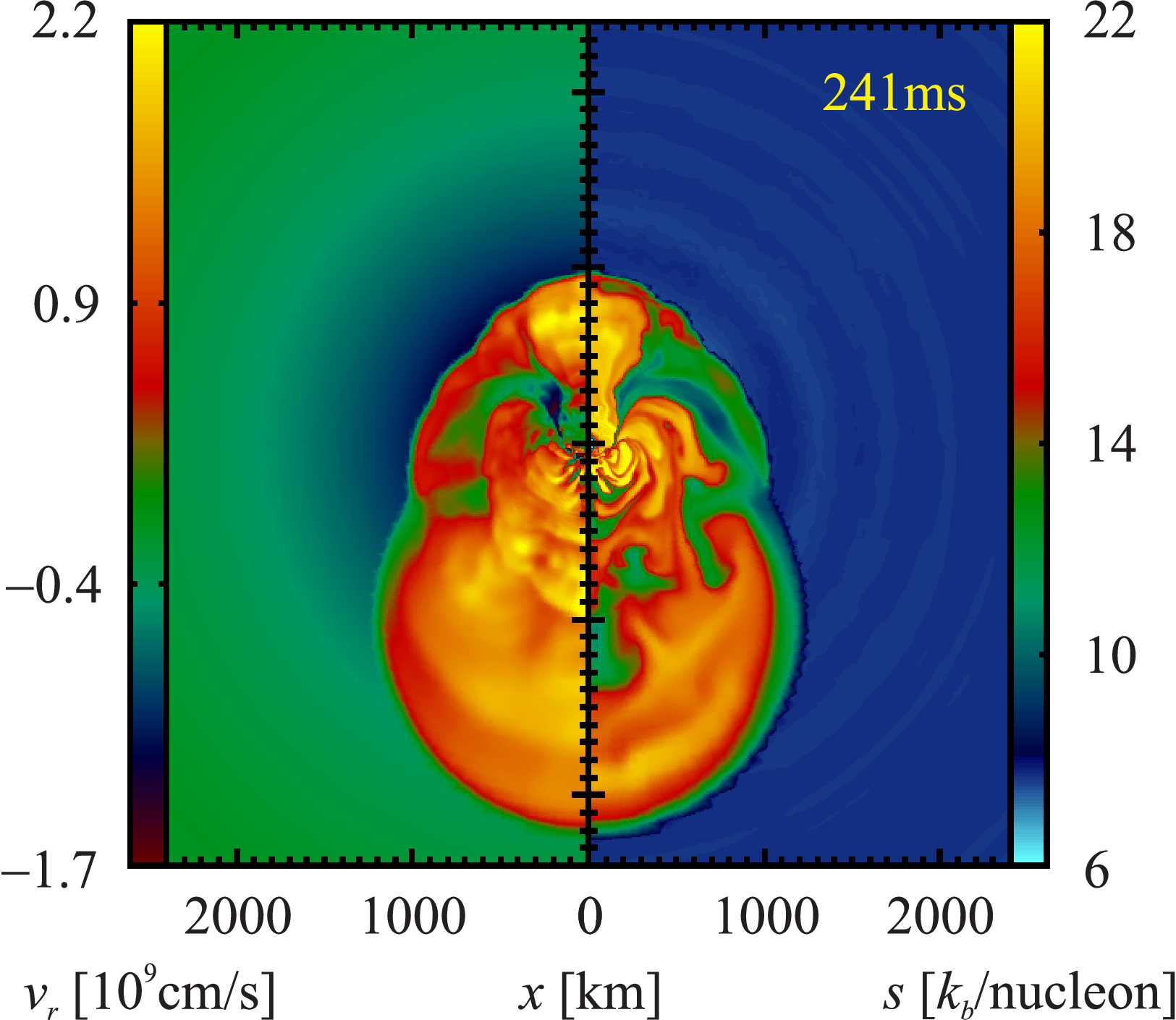}
\caption{Snapshots of the evolution of model u8.1, depicting the
  radial velocity $v_r$ (left half of panels) and the entropy per
  baryon $s$ (right half of panels) $66 \ \mathrm{ms}$, $71
  \ \mathrm{ms}$, $83 \ \mathrm{ms}$, $140 \ \mathrm{ms}$, $200
  \ \mathrm{ms}$, and $241 \ \mathrm{ms}$ after bounce (from top left
  to bottom right). Once the gain region becomes convectively
  unstable, small-scale plumes begin to grow ($66 \ \mathrm{ms}$, $71
  \ \mathrm{ms}$) and merge into somewhat larger structures on the
  scale of $10^\circ \ldots 50^\circ $($83 \ \mathrm{ms}$).  As
  convection becomes more violent, the deformation of the shock
  becomes more pronounced ($140 \ \mathrm{ms}$), and a dipolar
  asymmetry finally develops after the shock starts to accelerate
  outward ($201 \ \mathrm{ms}$, $241 \ \mathrm{ms}$). 
\label{fig:u81}
}
\end{figure*}

\begin{figure*}
\plottwo{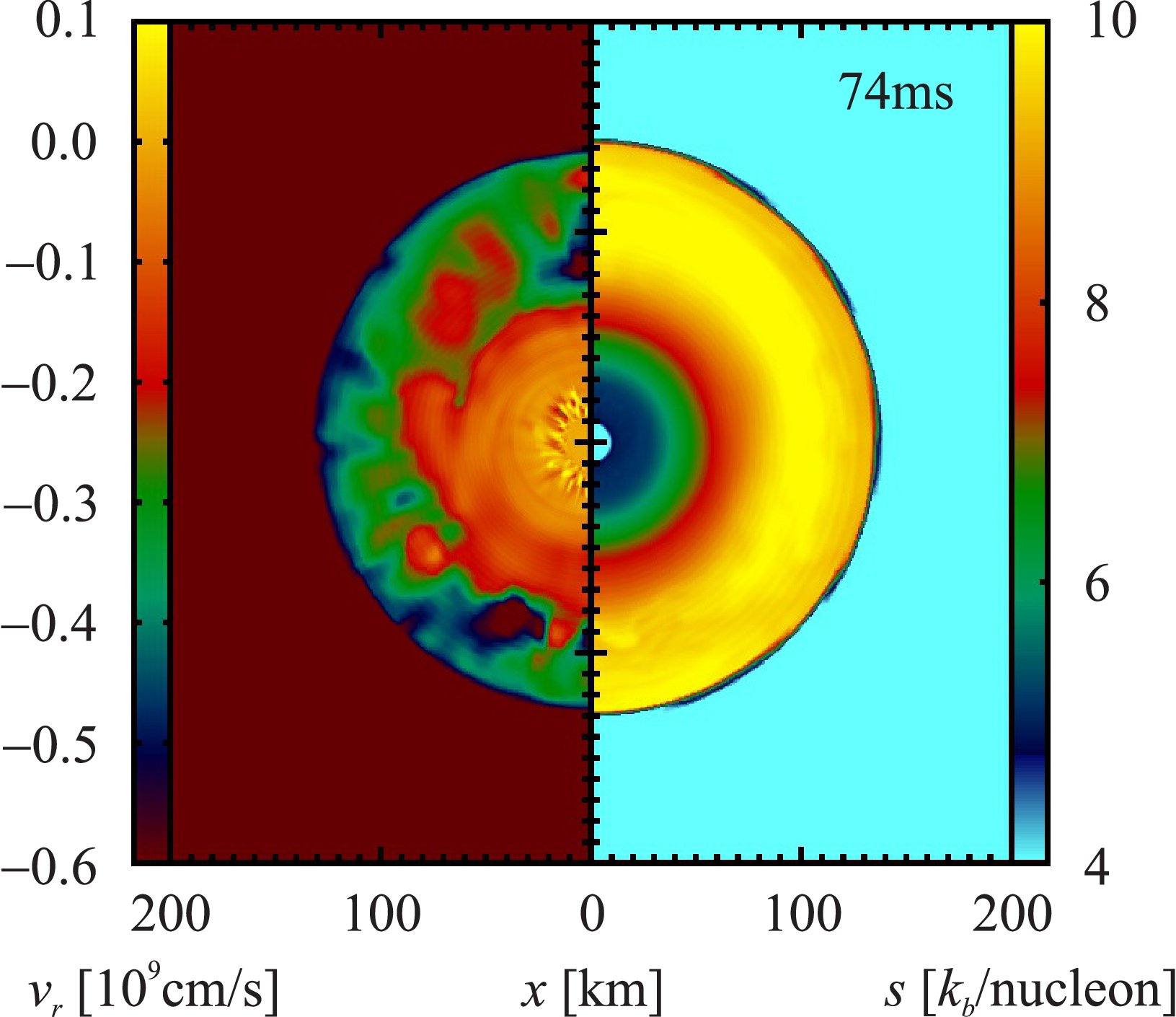}{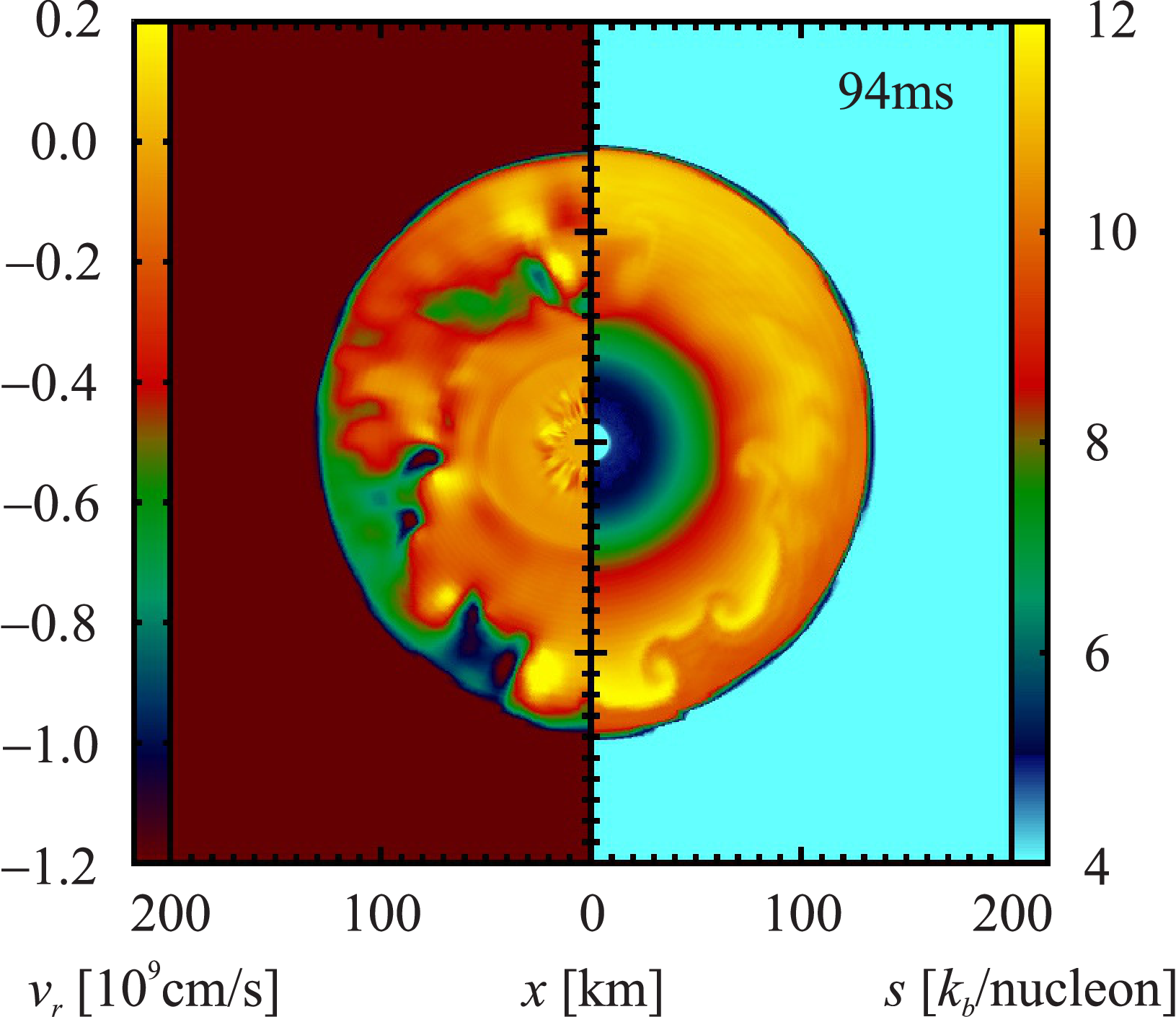}
\vspace{0.2in}
\plottwo{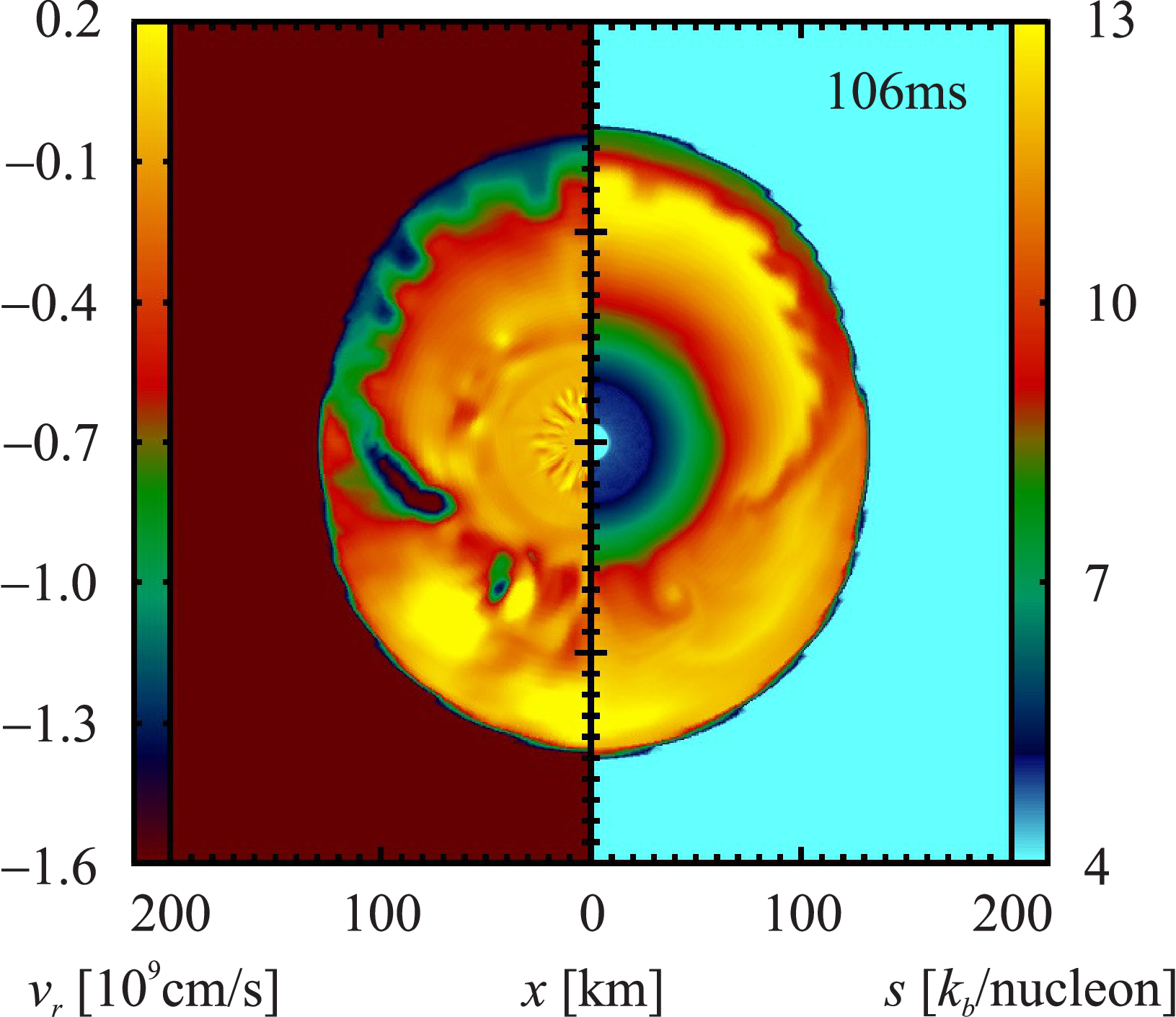}{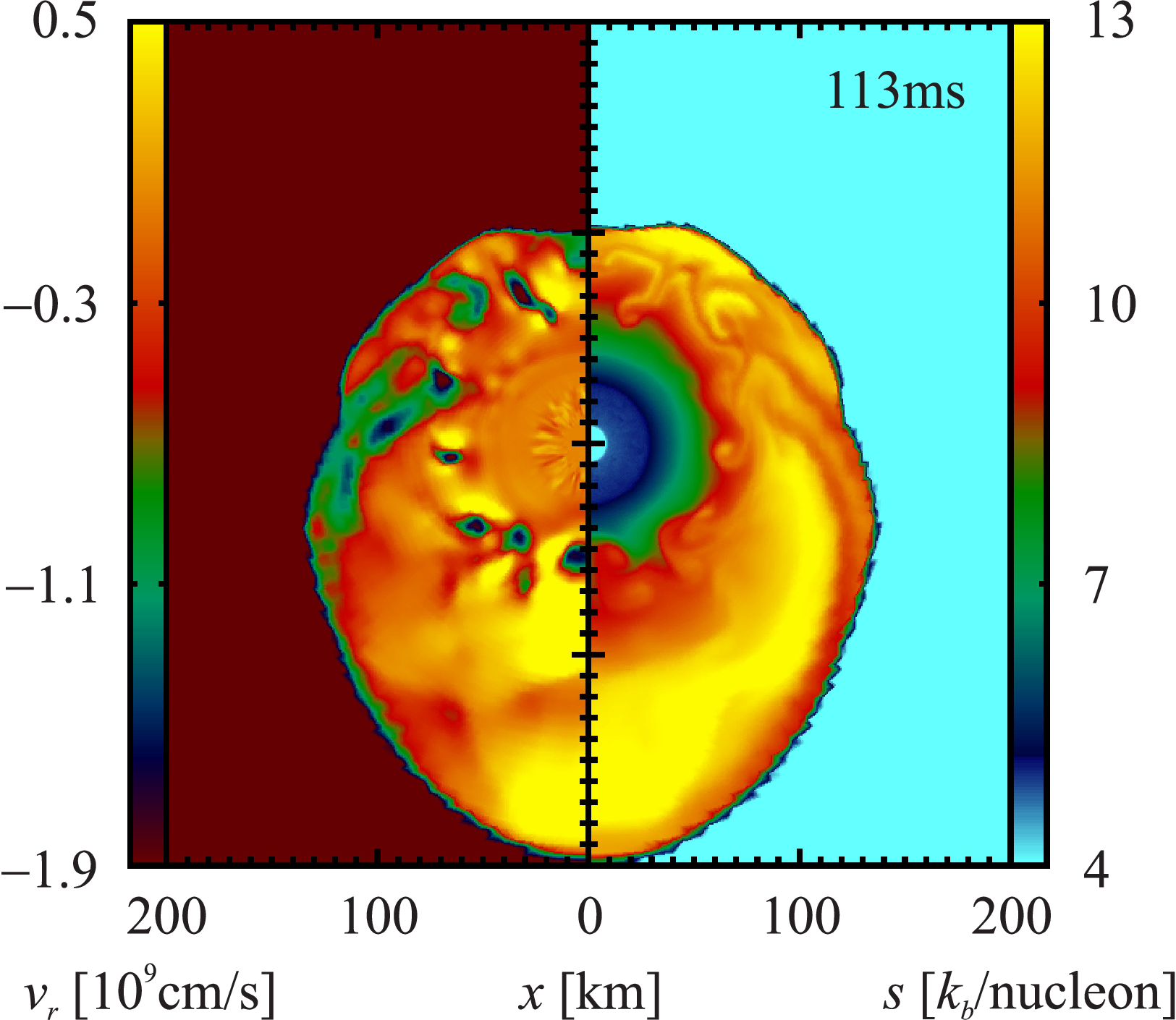}
\vspace{0.2in}
\plottwo{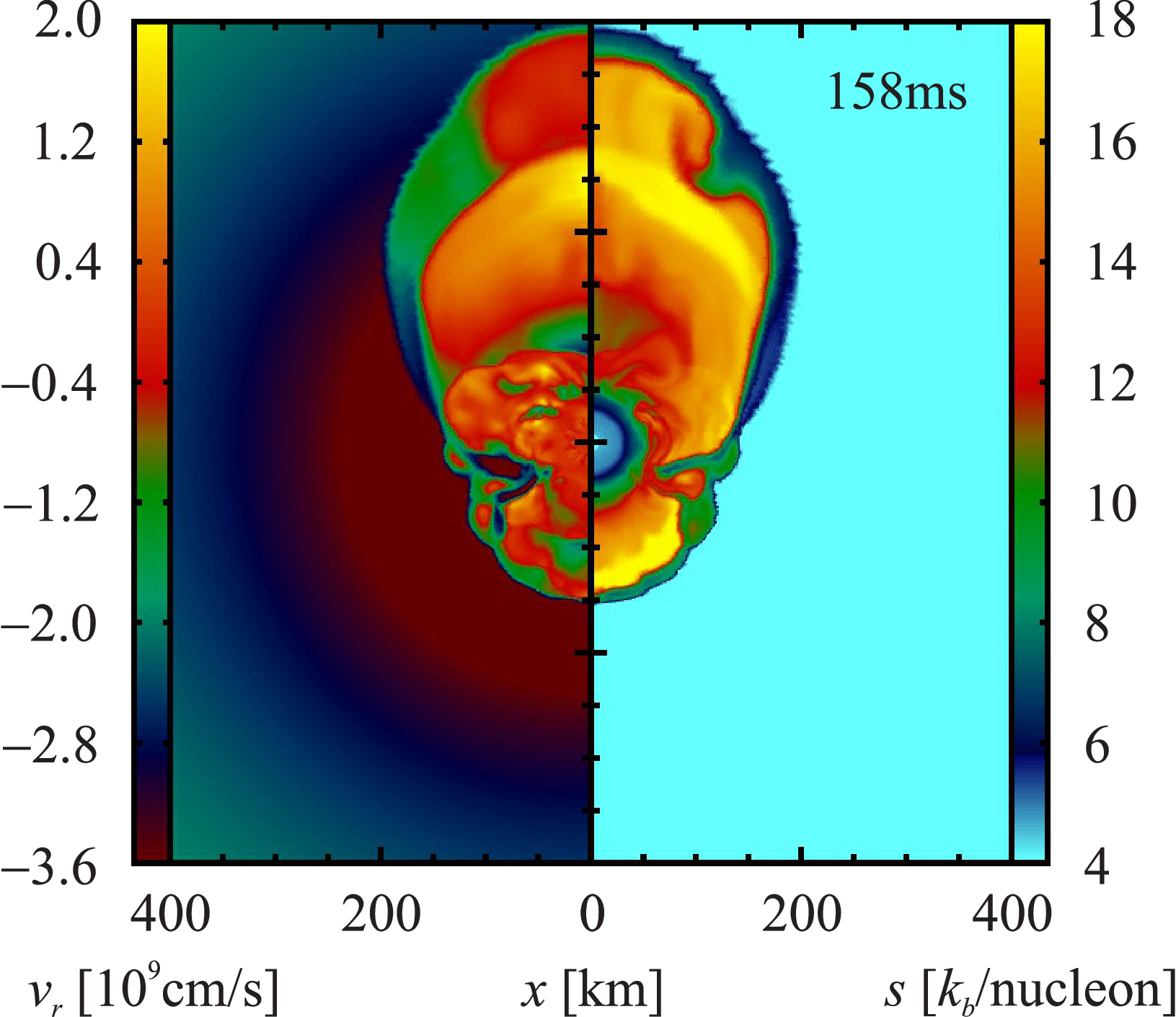}{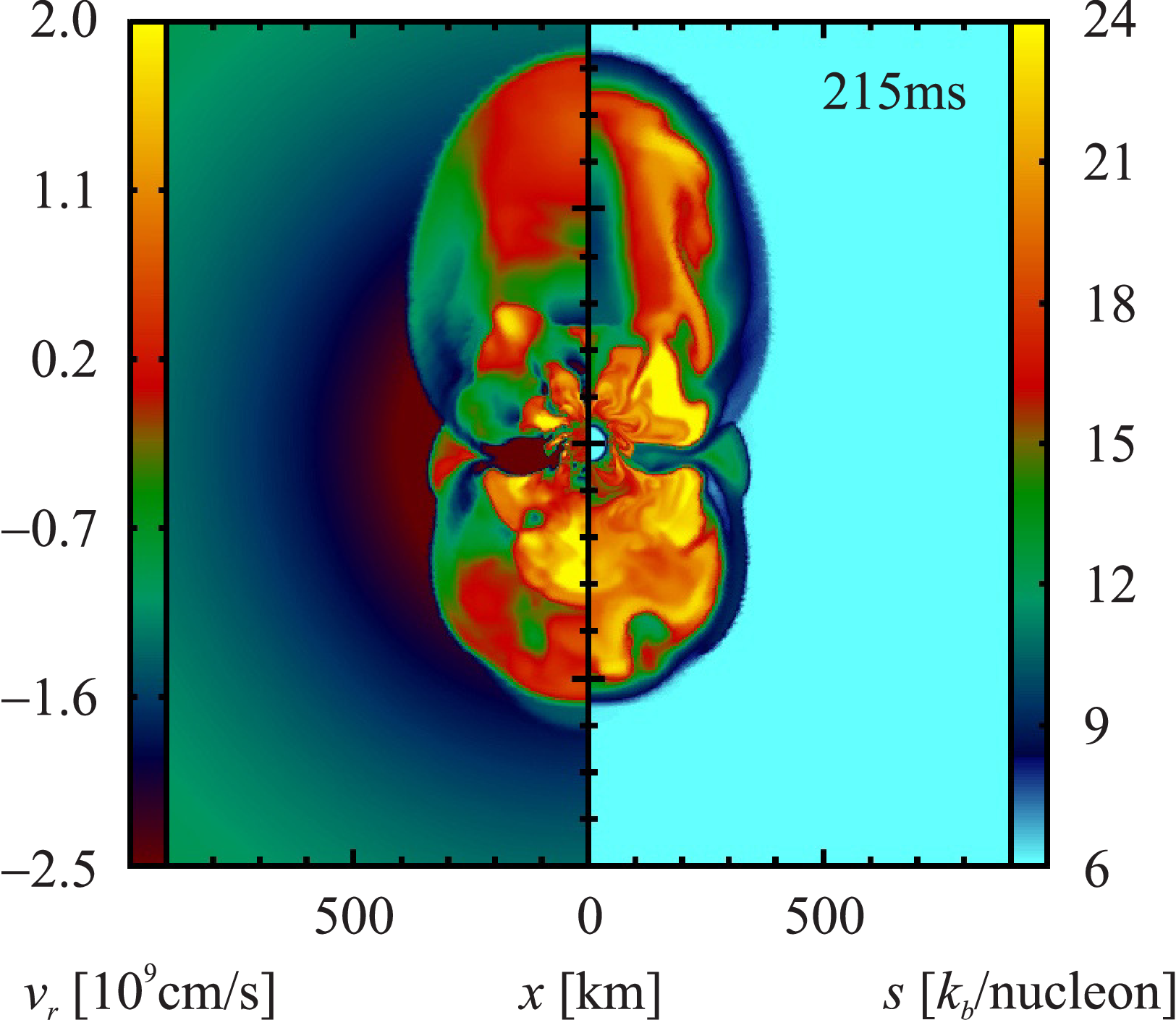}
\caption{Snapshots of the evolution of model s27.0, depicting the
  radial velocity $v_r$ (left half of panels) and the entropy per
  baryon $s$ (right half of panels) $74 \ \mathrm{ms}$, $94
  \ \mathrm{ms}$, $106 \ \mathrm{ms}$, $113 \ \mathrm{ms}$, $159
  \ \mathrm{ms}$, and $215 \ \mathrm{ms}$ after bounce (from top left
  to bottom right). The six panels exemplify the growth of the SASI in
  the linear regime ($74 \ \mathrm{ms}$), the development of parasitic
  instabilities on top of a clear $\ell=1$ asymmetry of the global
  entropy distribution ($94 \ \mathrm{ms}$, $106 \ \mathrm{ms}$), the
  transition to the non-linear regime with the formation of a
  pronounced downflow ($106 \ \mathrm{ms}$ to $159 \ \mathrm{ms}$) and
  the phase of vigorous shock expansion ($215 \ \mathrm{ms}$). Note
  the clearly different structure in the first four panels compared to
  the corresponding panels in Figure~\ref{fig:u81}.
\label{fig:s27}
}
\end{figure*}

\begin{figure*}
\plottwo{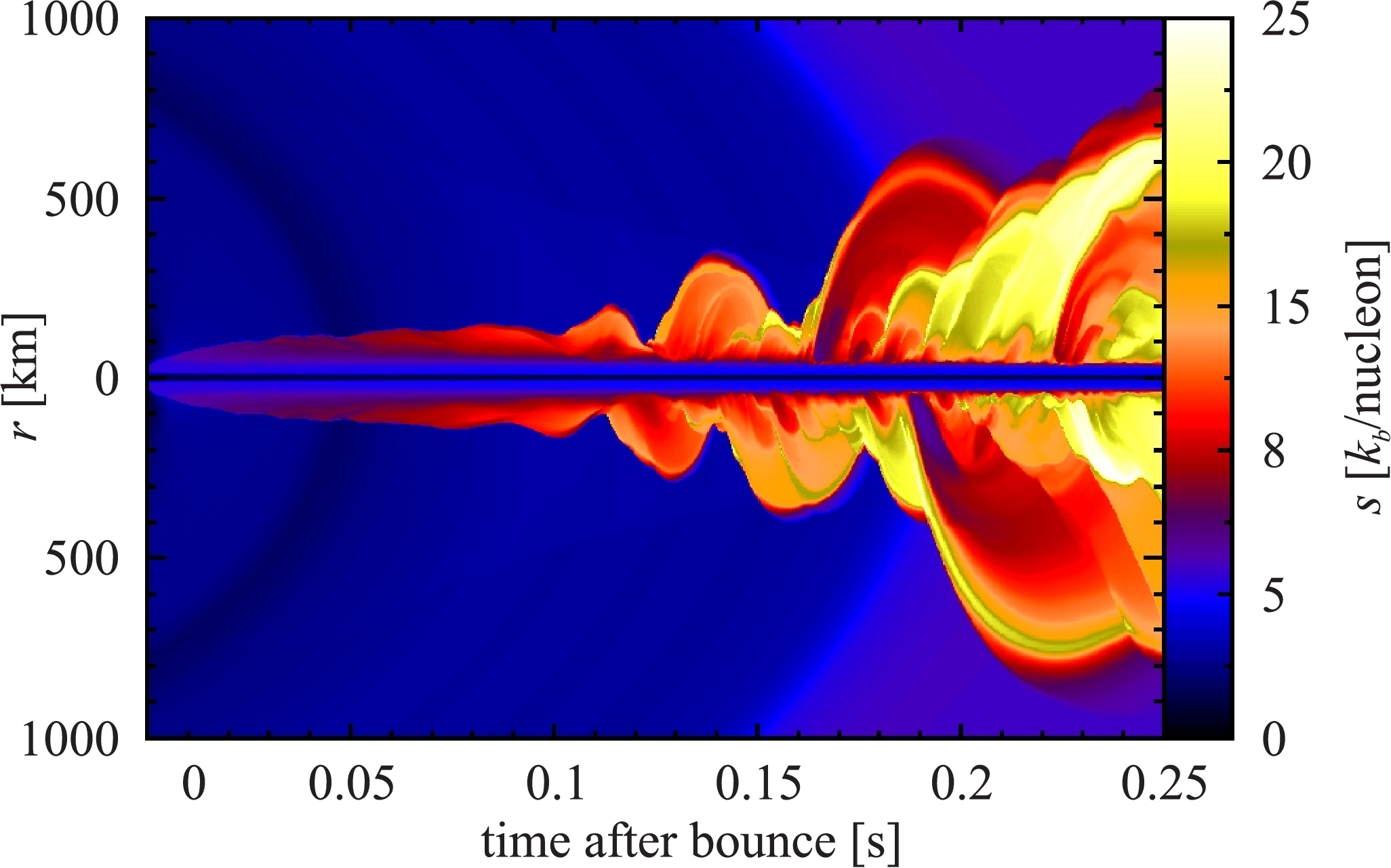}{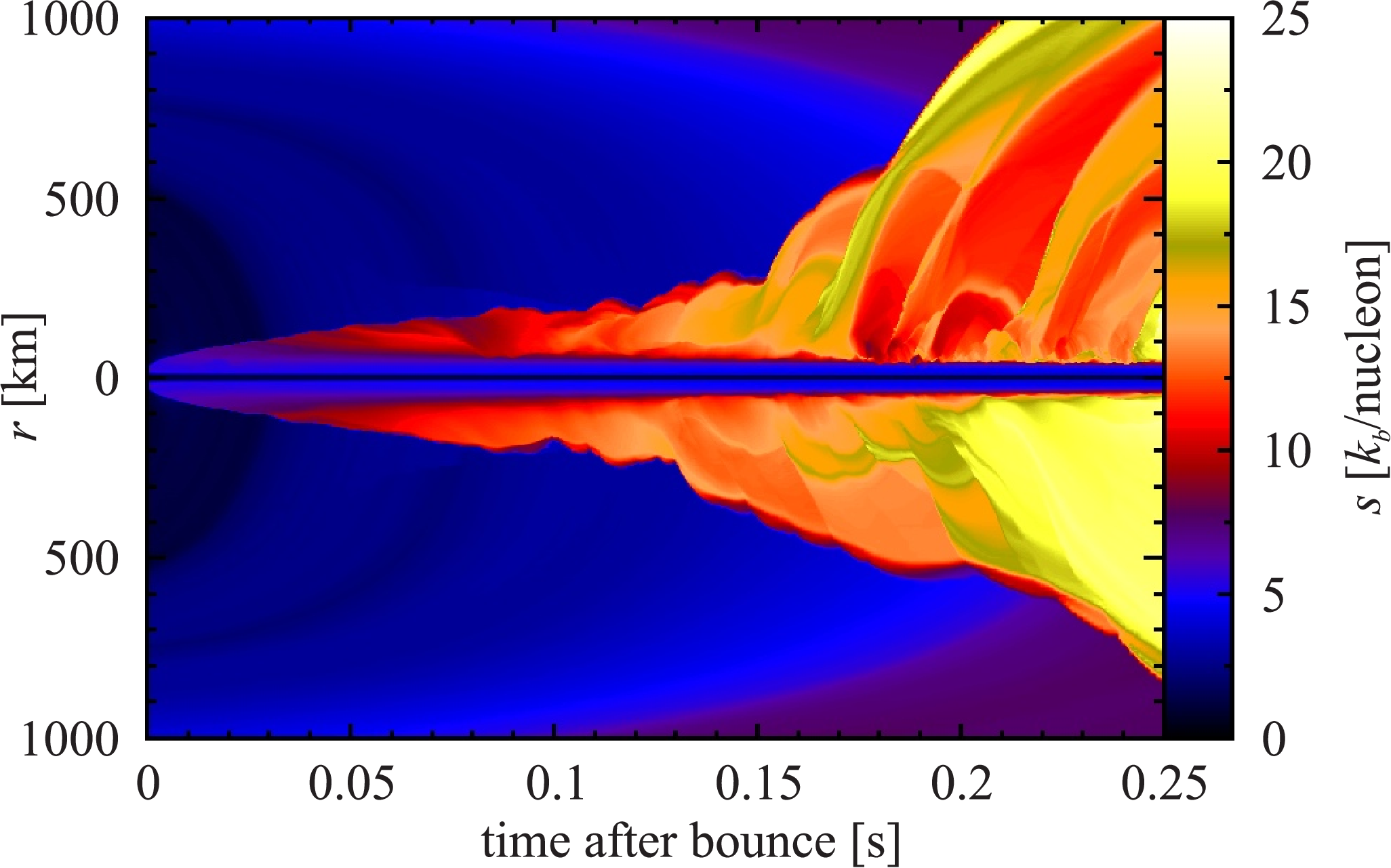}
\caption{Entropy along the north and south polar axis as a function of
  time for models s27.0 (left) and u8.1 (right).
\label{fig:entropy}
}
\end{figure*}

\begin{figure*}
\plottwo{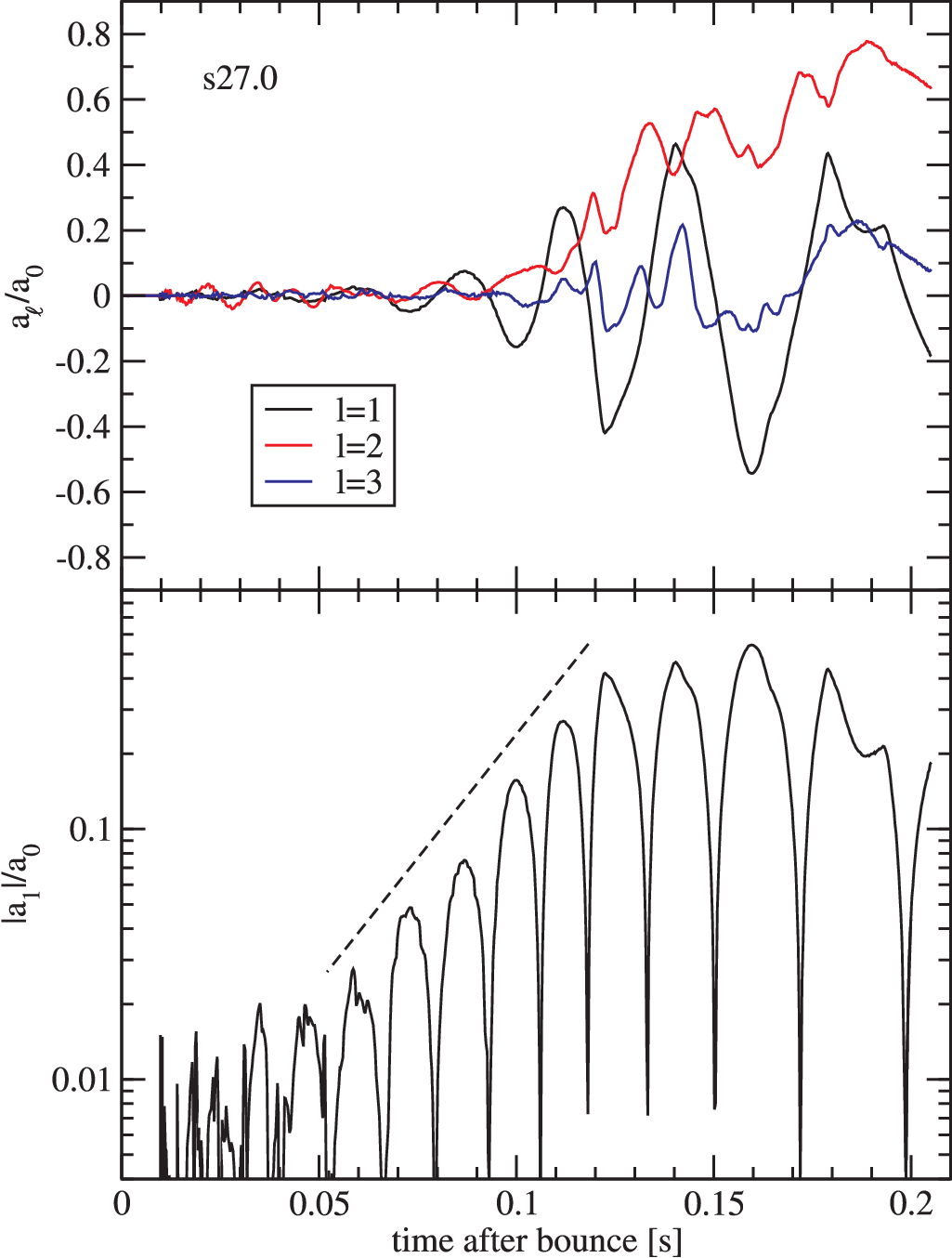}{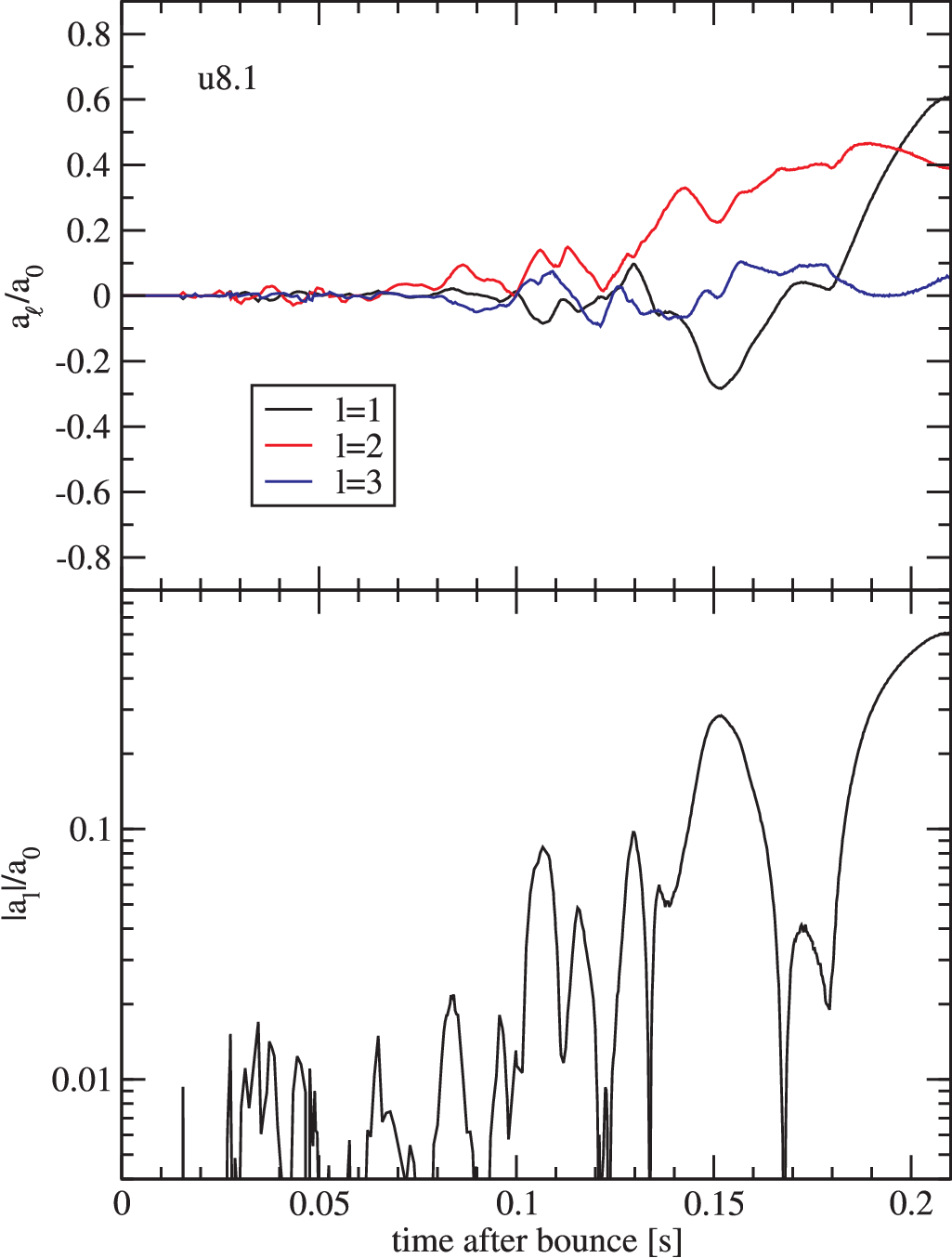}
\caption{Coefficients for the decomposition of the shock surface into
  Legendre polynomials. The top panels show the first three normalized
  Legendre coefficients $a_\ell/a_0$ for model s27.0 (left) and model
  u8.1 (right). In order to better exemplify the growth of the $\ell=1$
  mode, $\left| a_1 \right|/a_0$ is also plotted on a logarithmic
  scale in the bottom panels. The dashed line roughly denotes the
  slope of the exponential growth of the SASI up to $\sim \! 120
  \ \mathrm{ms}$ in model s27.0.
\label{fig:sasi}
}
\end{figure*}

Superficially, model s27.0 and u8.1 might appear to evolve in a very
similar fashion: Roughly around $120 \ \mathrm{ms}$ after bounce the
average shock radius starts to increase, and by $200 \ \mathrm{ms}$ the
shock is already expanding rapidly, although model s27.0 evidently lags
behind u8.1 a little (Figure~\ref{fig:shock}).  Especially during the
later phases, the shock becomes strongly deformed with a ratio
$r_\mathrm{max}/r_\mathrm{min}$ of the maximum and minimum shock
radius on the order of $2 \ldots 3$. Both models seems to provide
similar examples for an explosion at a relatively early stage.

However, this appearance is deceptive: A hint at more profound
differences between s27.0 and u8.1 is already furnished by the
critical ratio $\tau_\mathrm{adv}/\tau_\mathrm{heat}$ of the
``advection'' or ``residence'' time-scale and the heating time-scale
for the material in the gain region, which serves as an indicator for
an explosive runaway due to neutrino energy deposition (for
$\tau_\mathrm{adv}/\tau_\mathrm{heat}>1$; see
\citealt{janka_01,thompson_05,buras_06b,murphy_08,fernandez_12}). Figure~\ref{fig:timescale_ratio}
shows that model s27.0 approaches the critical threshold much later
than model u8.1, i.e.\ at roughly $\sim \! 180 \ \mathrm{ms}$ instead
of $\sim \! 110 \ \mathrm{ms}$. Nevertheless, even though the runaway
condition is not yet met, the shock already expands considerably
before that time in model s27.0. This suggests that at least for the first $\sim \! 180
\ \mathrm{ms}$ there may be a
driving agent other than neutrino heating that is responsible for
pushing the shock outwards.  One should bear in mind,
  however, that it is not completely clear for which value of
  $\tau_\mathrm{adv}/\tau_\mathrm{heat}$ one could already expect a
  noticeable expansion of the shock: Neutrino heating might drive
  considerable shock expansion even for
  $\tau_\mathrm{adv}/\tau_\mathrm{heat}<1$ depending on progenitor
  specifics.  However, it seems inevitable that large aspherical
  motions in the gain regions with Mach numbers on the order of $\sim
  1$ will affect the structure of the accretion flow, including the
  shock position (cp.\ Section~\ref{sec:growth}).

\subsection{Growth of Instabilities}
\label{sec:growth}

\begin{figure*}
\plottwo{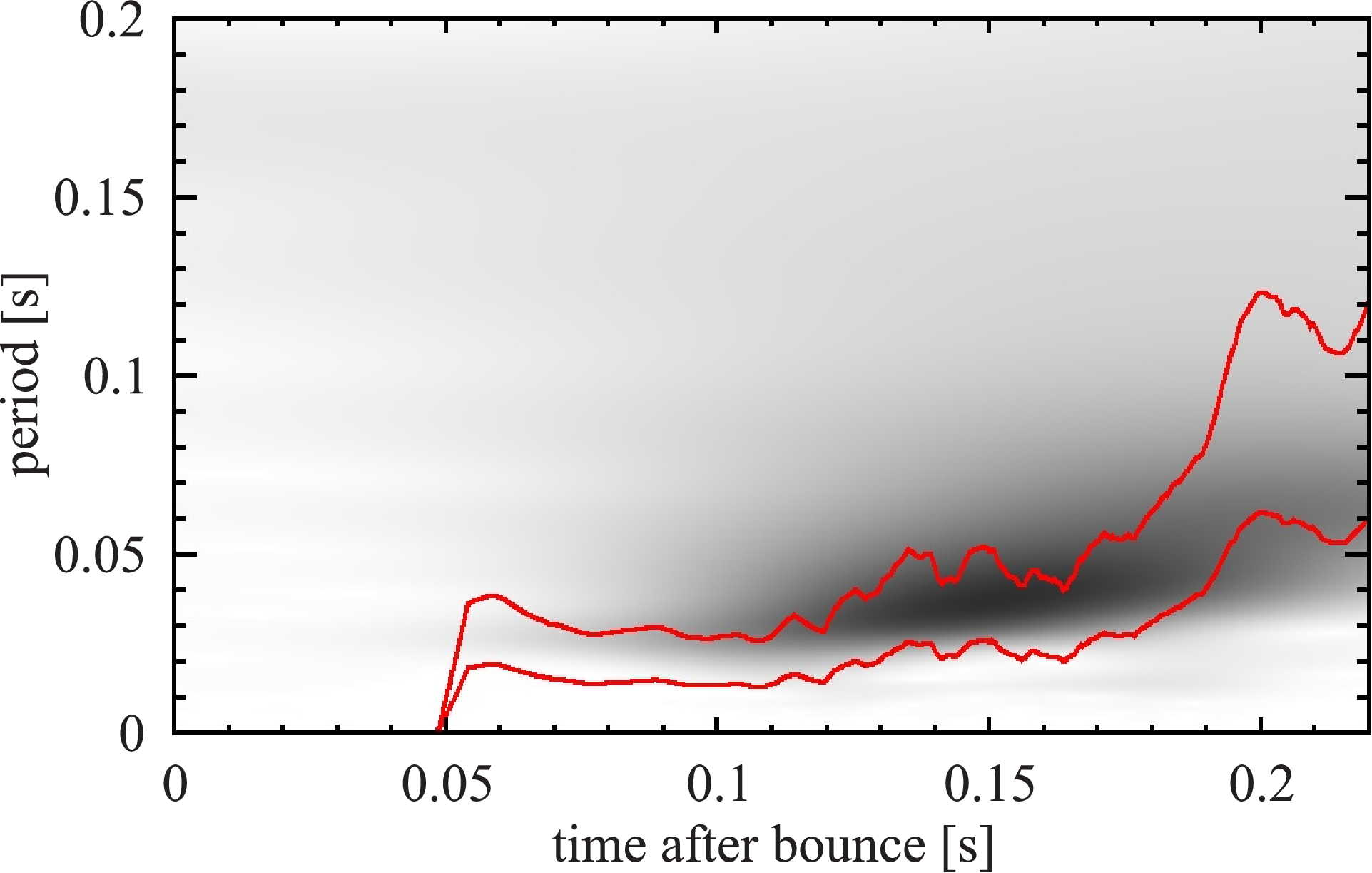}{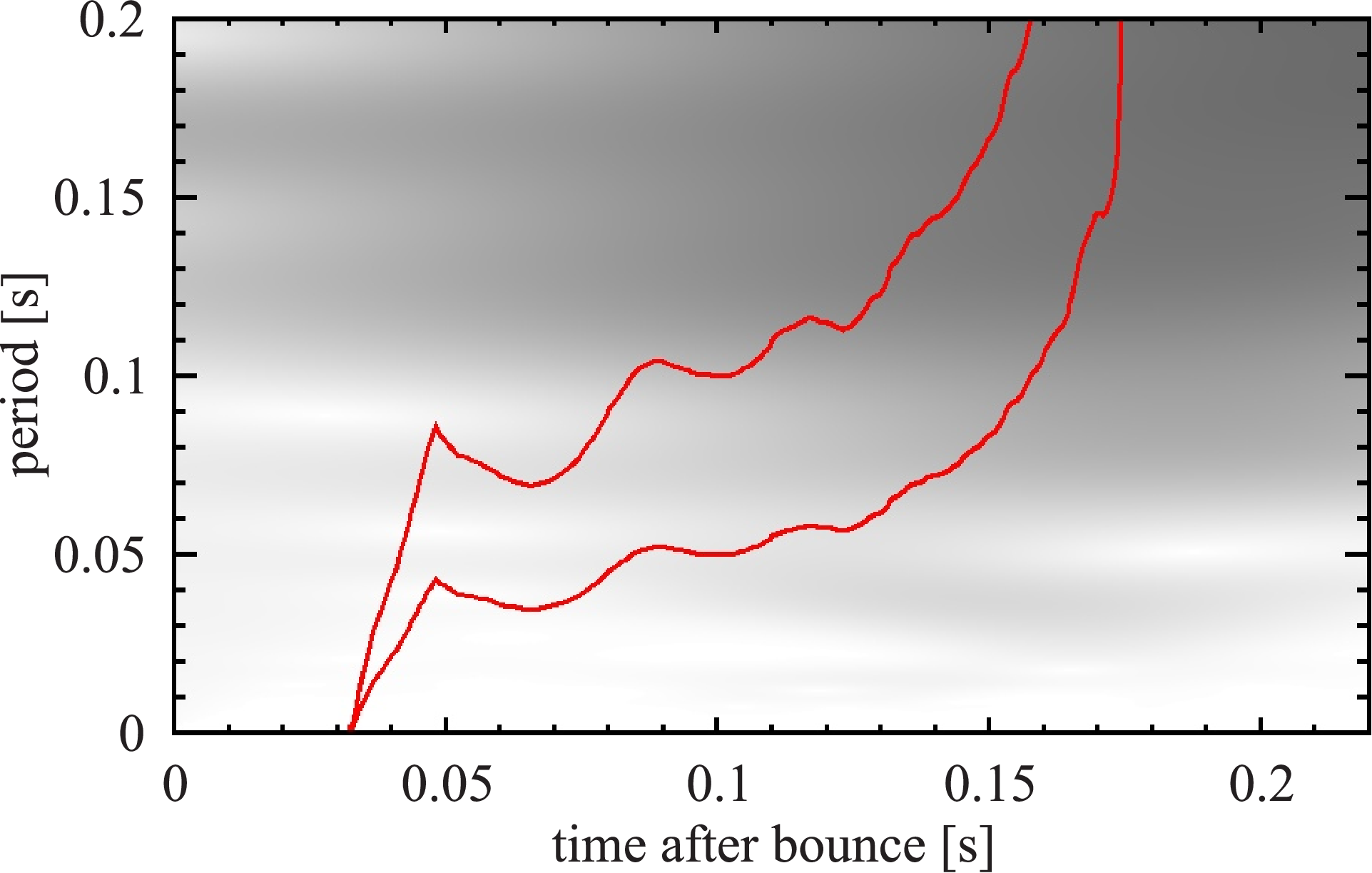}
\caption{Wavelet spectrograms of the normalized Legendre
  coefficient $a_1/a_0$ of the $l=1$ mode of the shock for model s27.0
  (left) and model u8.1 (right).  Large values of the wavelet
  transform are indicated by darker shades.  The red curves roughly
  define a band in which an advective-acoustic instability should
  surface; the upper and lower curves show the (smoothed) time
  evolution of the advection time-scale $\tau_\mathrm{adv}$ and of $2
  \tau_\mathrm{adv}$, respectively.
\label{fig:wavelet}
}
\end{figure*}

The reason for the peculiar evolution of the $27 M_\odot$ progenitor is
to be sought in the strong and relatively unimpeded growth of the SASI
as primary instability during the first $\sim \! 200 \ \mathrm{ms}$ as
opposed to neutrino-driven convection in the $8.1 M_\odot$ star -- a
feature hitherto not reported from full multi-group neutrino
hydrodynamics simulations \citep{marek_09,mueller_12}.

As shown by Figures~\ref{fig:s27} and \ref{fig:u81}, the morphology of
the post-shock flow in model u8.1 and model s27.0 becomes quite
dissimilar as soon as the gain region forms a few tens of milliseconds
after bounce. Model u8.1 (Figure~\ref{fig:u81}) conforms to the
``standard'' behavior exhibited by all recent first-principle
explosion simulations \citep{marek_09,mueller_12}: The convective
instability develops first, and perturbations at intermediate scales
(corresponding to multipole orders of $\ell \gtrsim 5\ldots 10$) grow
fastest, cf.\ \citealp{foglizzo_06,foglizzo_07}). Larger plumes only
form once convection has become vigorous and the shock has already
expanded quite considerably (cp.\ the $11.2 M_\odot$ case of
\citealt{mueller_12}). A strong non-oscillatory dipolar shock
deformation appears at late time once the explosion is underway, and
this deformation seems to be driven by buoyant neutrino-heated
plumes rather than by genuine SASI activity.

Model s27.0 behaves in a completely different manner
(Figure~\ref{fig:s27}).  Here we first observe small, but clearly
recognizable linear SASI oscillations (first panel of
Figure~\ref{fig:s27}). Later, as the amplitude of the SASI increases,
parasitic Rayleigh-Taylor and Kelvin-Helmholtz instabilities grow on
top of the SASI flow, but these remain localized, appearing and
disappearing as the shock oscillates back and forth, and never growing
into developed convection. The different character of the
  post-shock flow can also be seen in Figure~\ref{fig:entropy}, which
  shows the time-evolution of the entropy along the polar axis for
  both models and illustrates that -- unlike model 8.1, s27.0 indeed
  exhibits the sloshing motions characteristic of the SASI. Until $\sim
  \! 180 \ \mathrm{ms}$ the expansion and contraction of the shock in
  the northern and southern hemispheres are nicely antisynchronized.

The velocity fluctuations in the gain region steadily grow in
magnitude in model s27.0, eventually leading to the
formation of secondary shocks (see the fifth panel of
Figure~\ref{fig:s27}). It is likely that these strong velocity
fluctuations
%, which reach the order of the speed of sound, 
are responsible for pushing the shock
further out because they provide an additional Reynolds stress
contribution for the angle-averaged and time-averaged flow on top of
the thermal pressure. Moreover, the thermal pressure itself is
increased once dissipation in secondary shock starts to convert
kinetic energy stored in the SASI motions into thermal energy.
The dynamical importance of both these effects depends
on the typical velocity $\delta v$ of the aspherical flow perturbations,
which determines the ratio of the Reynolds stresses $\rho\, \delta v^2$
and the kinetic energy to the thermal pressure $P_\mathrm{therm}$ and energy
density $\rho \epsilon_\mathrm{therm}$, respectively:
\begin{equation}
\rho \,\delta v^2/P_\mathrm{therm} \propto \left( 1/2 \rho \, \delta v^2\right)/ \left(\rho \epsilon_\mathrm{therm}\right) \propto \delta v^2/c_s^2 = \mathrm{Ma}^2
\end{equation}
Once the SASI reaches the non-linear stage in model s27.0, the
typical Mach number $\mathrm{Ma}$ (determined from angular root
mean square averages of the lateral velocity component) exceeds
$\gtrsim 0.5$ in large parts of the gain region and peaks
at $\sim 1$ near the average shock radius, clearly indicating
the dynamical importance of aspherical instabilities in
that model. In that sense, model s27.0 may provide an example for
  SASI-driven shock expansion \emph{prior} to the explosion as opposed to
  the neutrino-driven shock expansion that roughly sets in as
  $\tau_\mathrm{adv}/\tau_\mathrm{heat}$ exceeds unity. 
 
We emphasize that these mechanisms for shock expansion
  are, of course, not specific to the SASI; indeed \citet{murphy_12}
  demonstrated that the inclusion of the Reynolds stresses in the
  shock jump conditions largely accounts for the increased shock radii
  in their convective models. However, while the immediate cause for
  the expansion of the shock is similar in model s27.0 to
  convectively-aided explosion models like u8.1 or those of
  \citet{murphy_12}, the ultimate source for the energy of aspherical
  motions is different: As long as the SASI remains the dominant
  instability, it essentially feeds directly on the kinetic energy of
  the accreted matter whereas convection is (indirectly) powered by
  neutrino heating. 

\begin{figure*}
\plottwo{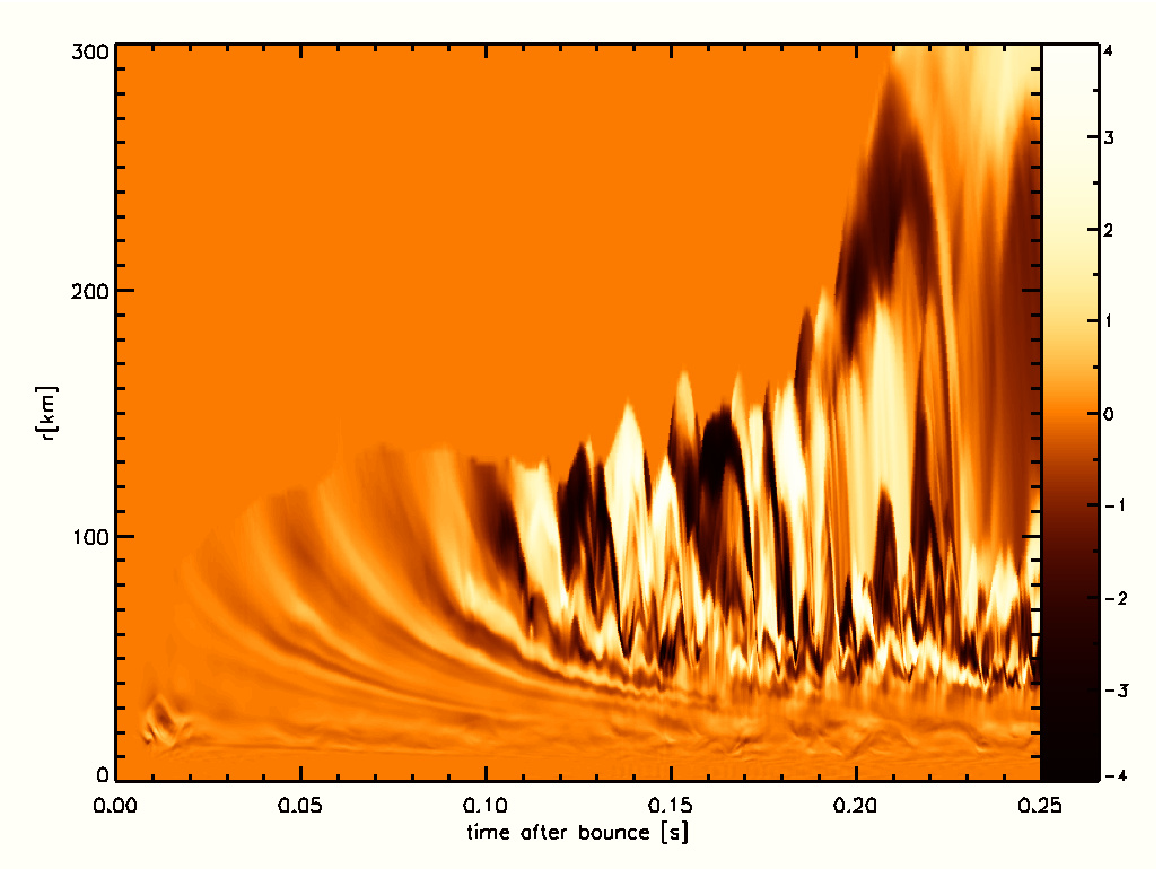}{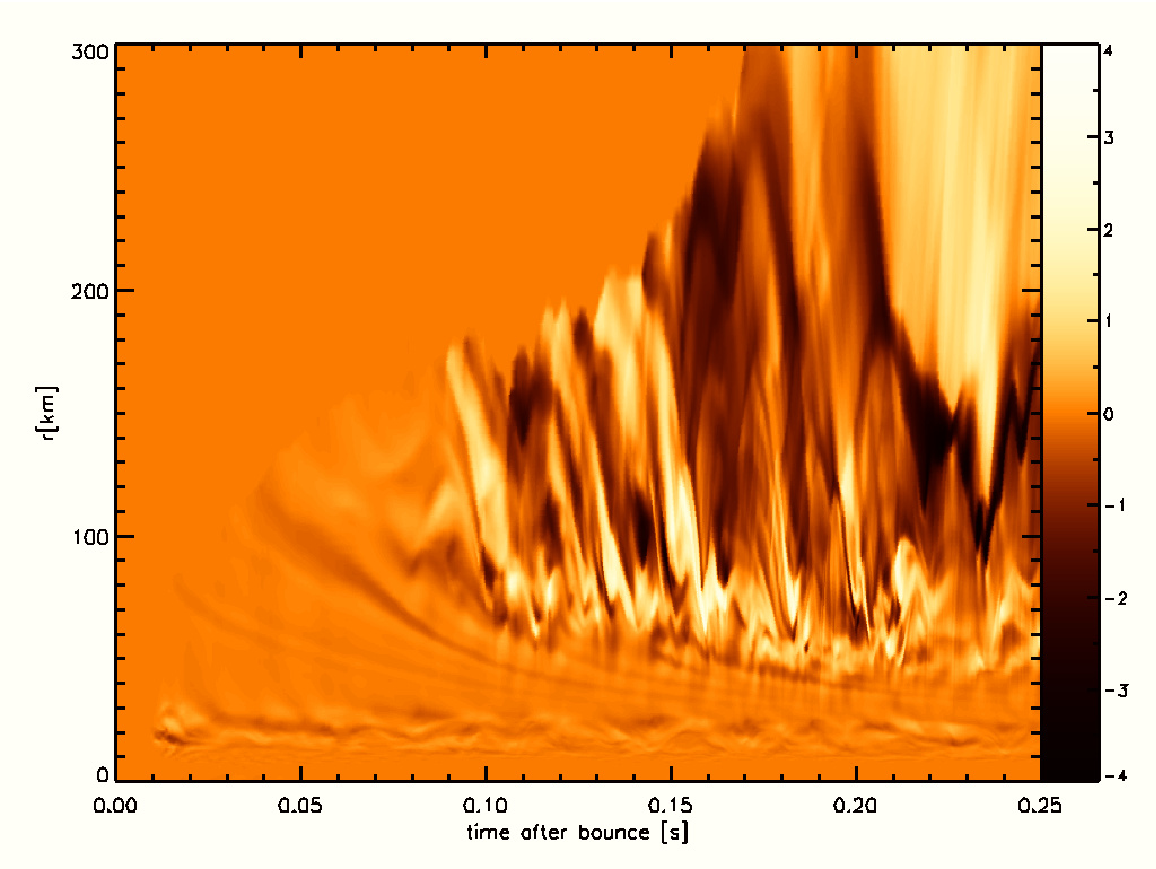}
\caption{Lateral velocity $v_\theta$ in the 
equatorial plane as a function of time and radius for
model s27.0 (left) and model u8.1 (right).
\label{fig:lateral_velocity}
}
\end{figure*}

\begin{figure}
\plotone{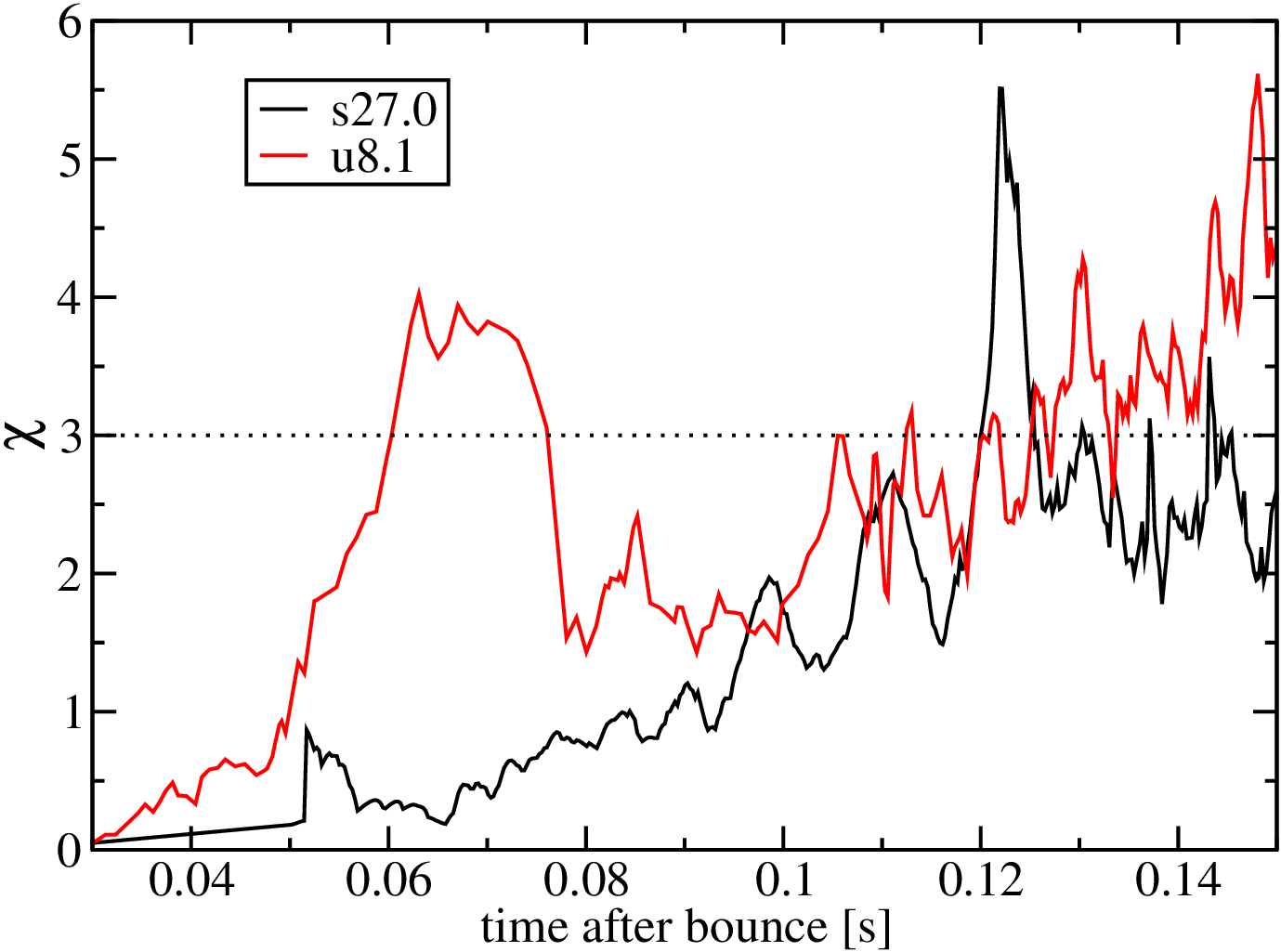}
\caption{Evolution of the stability parameter $\chi$ for model s27.0
  (black) and model u8.1 (red). The threshold value of $\chi \approx
  3$ for convection is reached about $60 \ \mathrm{ms}$ after bounce
  for model u8.1, whereas convective stability applies for s27.0 until
  $120 \ \mathrm{ms}$ when the SASI has already grown to the
  non-linear regime. Note that $\chi$ becomes less meaningful as the
  post-shock flow develops strong asphericities. 
For this reason, e.g., the fact that $\chi$ drops
below three at $t \approx 0.08 \ \mathrm{s}$ once convection
has become vigorous does not indicate declining convective activity.
\label{fig:chi}}
\end{figure}

\begin{figure}
\plotone{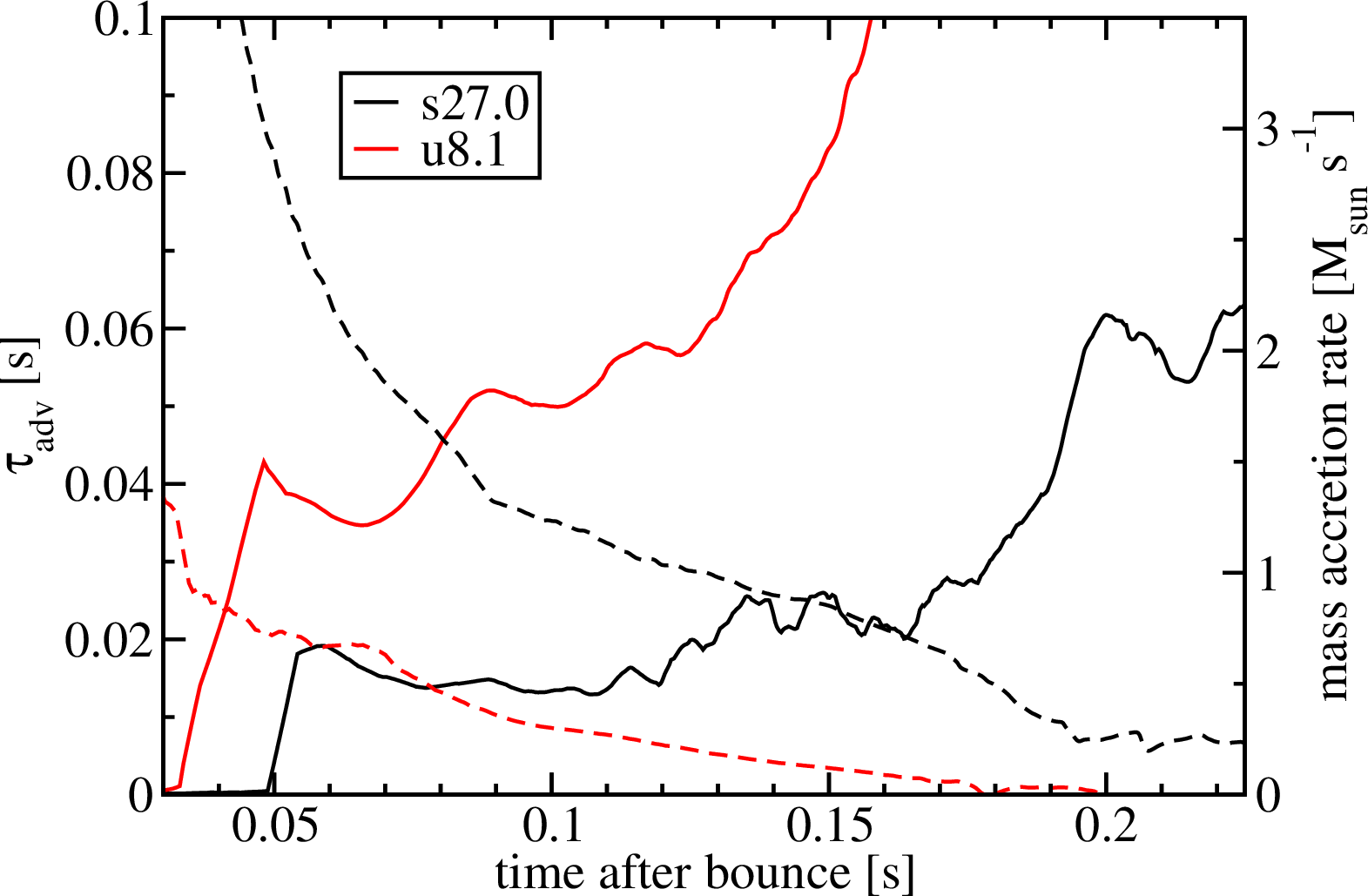}
\caption{Advection time-scale $\tau_\mathrm{adv}$ (solid lines) and
  mass accretion rate $\dot{M}$ (dashed lines) for model s27.0 (black)
  and model u8.1 (red). The time-scales are evaluated as described in
  \citet{mueller_12}, and the curves have been smoothed using a
  running average over $5 \ \mathrm{ms}$.
\label{fig:tadv}
}
\end{figure}

\subsection{Evidence for the SASI -- Quantitative Analysis}
For model s27, we thus have a SASI-like flow
  morphology reminiscent of simulations with no \citep{blondin_03} or
  suppressed \citep{scheck_08} neutrino heating. The
associated aspherical velocities apparently become large enough (i.e.\ on the
order of the speed of sound) to be dynamically relevant. But model
s27 furnishes further evidence pointing to the SASI as the
initially dominant instability besides the qualitative flow
morphology.

A mode analysis of the shock surface, based on a decomposition into
Legendre polynomials, confirms the dramatic differences in the growth
of hydrodynamical instabilities (Figure~\ref{fig:sasi}). The expansion coefficients $a_\ell$
are given by
\begin{equation}
a_\ell=
\frac{2 \ell + 1}{2}
\int\limits_{0}^\pi r_\mathrm{sh}(\theta) P_\ell (\cos \theta) \sin \theta
\, \mathrm{d} \theta,
\end{equation}
and the first three coefficients $a_1$, $a_2$, and $a_3$ (normalized
to the angle-averaged shock position $\left\langle r_\mathrm{sh}
\right\rangle = a_0$). The convectively-dominated model u8.1 
(right panel of Figure~\ref{fig:sasi}) is
characterized by relatively weak shock oscillations until $\sim \! 130
\ \mathrm{ms}$ when the explosive runaway due to neutrino heating is
already underway.  Afterwards, a pronounced dipole ($\ell=1$) and
quadrupole ($\ell=2$) mode appear, but there is no trace of periodic
oscillations at any stage. Apparently, the growth of the low-$\ell$ modes
is a stochastic process driven by the convective plumes in this case.
Later in the explosion, the geometry becomes more stable with a big
buoyant bubble in the southern hemisphere. The dominance of such
large-scale bubbles at late times is in agreement with analytic
estimates for extended convective regions with a small, almost
point-like heating source \citep{chandrasekhar_61,foglizzo_06}.

By contrast, model s27.0 (left panel of Figure~\ref{fig:sasi})
exhibits shock oscillations with a well-defined periodicity for the
$\ell=1$ mode as expected for a relatively unperturbed SASI (cf.\ the
numerical experiments by \citealt{scheck_08}). The expected
exponential growth of the dipole amplitude $| a_1 |/a_0$ from $\sim \!
0.02$ to $\sim \! 0.4$ during the linear phase can be observed for
several oscillation periods, and a growth rate $\omega_\mathrm{SASI}
\approx 45 \ \mathrm{s}^{-1}$ can be deduced. The quadrupole amplitude
($\ell=2$) increases in a non-oscillatory manner after $\sim \! 90
\ \mathrm{ms}$, which may be a result of the transition to the
non-linear regime.  Moreover, we expect the
  oscillation period to scale with the advection time-scale by a
  factor between unity and $\lesssim 2$ if an advectice-acoustic cycle
  operates in model s27.0 \citep{foglizzo_07,scheck_08,guilet_12}. This confirmed by a wavelet analysis of the
  $\ell=1$ coefficient of the shock position using the Morlet wavelet
  \citep{torrence_98} shown in Figure~\ref{fig:wavelet}.  Even after
  $\sim 150 \ \mathrm{ms}$, when the underlying linear SASI eigenmode
  can no longer be easily discerned in the post-shock flow (see the
  last two panels in Figure~\ref{fig:s27} and \ref{fig:u81}), the
  advection time-scale apparently still sets the oscillation
  period. This is in stark contrast to model u8.1, for which the
  wavelet spectrogram never shows a lot of power in the expected
  ``SASI frequency band''.  The antisynchronized variation of the
  shock radius along the north and south polar axis (left panel of
  Figure~\ref{fig:entropy}) that can be discerned until $\sim \! 180
  \ \mathrm{ms}$ also points to a global low-$\ell$ instability; such
  a correlated variation of the shock radius in opposite directions is
  probably hard to reconcile with a stochastic forcing of low-$\ell$
  modes by convection.  

Up to $\sim 150 \ \mathrm{ms}$, a closer look at the
  post-shock velocity field even reveals the advective part of the
  SASI amplification directly. Figure~\ref{fig:lateral_velocity}
  clearly shows the advection of coherent perturbations in the lateral
  velocity (consistent with the $\ell=1$ sloshing mode) from the shock
  to the deceleration region at the proto-neutron star surface for
  about $100 \ \mathrm{ms}$. Around $ \sim 150 \ \mathrm{ms}$, the
  perturbations become somewhat crinkled due to the action of
  parasitic instabilities, but a temporal quasi-periodicity can still
  be recognized.  In model u8.1, we see only very faint traces of such
  coherent perturbations prior to the onset of convection.

There are thus good reasons for classifying model
  s27 as ``SASI-dominated'' and model u8.1 as 
  ``convectivion--dominated'' at least for first $\sim 150
  \ \mathrm{ms}$. The late-time behavior around and after shock
  revival is probably more complicated in both cases. During this
  phase, neutrino heating evidently plays a major role for the
  dynamics and ultimately powers the explosion, and buoyancy-driven
  instabilities certainly become important in the process. However,
  this does not imply that steady-state convection alone furnishes an
  appropriate picture for the post-shock flow at this stage. The
  simulations rather suggest a complicated interplay between
  neutrino-heated, buoyancy-driven bubbles, internal shocks and
  acoustic waves, supersonic downflows, and an aspherical shock that
  creates large non-radial velocities in the post-shock region (bottom
  panels of Figures~\ref{fig:u81} and \ref{fig:s27}). The
  characterization of this phase will require a better understanding
  of the interaction of the different hydrodynamical instabilities in
  the fully non-linear phase.

\subsection{Conditions for SASI-Dominated and
Convectively-Dominated Flow}

Why are the two models s27.0 and u8.1 dominated by different
hydrodynamical instabilities, and why, in particular, is convection
suppressed for the $27 M_\odot$ progenitor despite neutrino heating in
the gain layer? The key for understanding the differences is connected
to the fact that the presence of a negative entropy gradient in the
heating region is not sufficient for the onset of convection as
pointed out by \citet{foglizzo_06}. Under the assumption of a stalled
accretion shock, any convective perturbation is advected out of the
gain layer within a finite time, and convection can develop only if
the perturbation is amplified sufficiently within this time
frame. Using a more rigorous mathematical analysis,
\citet{foglizzo_06} derived a growth parameter $\chi$ that is defined
in terms of the Brunt-{V\"ais\"al\"a} frequency $\omega_\mathrm{BV}$
and the spherically averaged advection velocity $\langle v_r\rangle$
as follows,
\begin{equation}
\label{eq:chi}
\chi = \int_{r_\mathrm{g}}^{\left \langle r_\mathrm{sh} \right \rangle} \frac{ \mathrm{Im}\, \omega_\mathrm{BV}}{\left| \langle v_r  \rangle \right|} \ud r,
\end{equation}
where the integral runs from the gain radius $r_\mathrm{g}$ to the
average shock radius $r_\mathrm{sh}$. Note that only the region where
$\omega_\mathrm{BV}^2<0$ indicates \emph{local} instability in the
fluid frame contributes to the integral. \citet{foglizzo_06} find a
threshold condition of $\chi \gtrsim 3$ for convective instability in
the gain region, which has been confirmed by parametrized as well as
first-principle simulations \citep{buras_06b,scheck_08,fernandez_09a}.
We emphasize that the criterion $\chi \gtrsim 3$ 
has been derived by \citet{foglizzo_06} for the \emph{linear} regime
and may no longer be useful for determining the presence
or absence of convective instability once significant
non-spherical perturbations develop in the post-shock region.

The time evolution of $\chi$ for our two models is shown in
Figure~\ref{fig:chi}. Clearly, the parameter $\chi$ indicates that
convection should set in at $\sim \! 60 \ \mathrm{ms}$ for u8.1, which
is exactly what we observe (Figure~\ref{fig:u81}). For s27.0, on the
other hand, $\chi$ is very low at this junction, and only reaches the
critical value $\chi=3$ some $120 \ \mathrm{ms}$ after bounce, when
the flow is already strongly aspherical due to the activity of the
SASI (Figure~\ref{fig:sasi}) and $\chi$ may no longer be a reliable
measure for convective instability.

The conditions for effective SASI growth are quite different.  Here, a
short advection time-scale $\tau_\mathrm{adv}$ 
is helpful: In the picture of the advective-acoustic cycle, the
linear growth rate $\omega_\mathrm{SASI}$ is given in terms
of the cycle efficiency $\mathcal{Q}$ and the duration $\tau_\mathrm{cyc}$
of the cycle by
\begin{equation}
\label{eq:omega}
\omega_\mathrm{SASI}=\frac{\ln |\mathcal{Q}|}{\tau_\mathrm{cyc}},
\end{equation}
as shown by \citet{foglizzo_06}. A smaller shock stagnation radius and
advection time-scale also imply a shorter cycle duration
$\tau_\mathrm{cyc}$, and hence conditions favorable for a more rapid
growth of the SASI.\footnote{Qualitatively, the same dependence would
  be expected for a purely acoustic cycle. Whether the SASI grows due
  to advective-acoustic or purely acoustic feedback is therefore
  irrelevant for our discussion.} Since $\tau_\mathrm{adv}$ is shorter
by a factor of $2 \ldots 4$ in model s27.0 compared to u8.1
(Figure~\ref{fig:tadv}), considerably more vigorous SASI activity is
to be expected. Moreover, convection could possibly destroy the
coherence of the entropy, vorticity, and acoustic waves involved in
the amplification cycle \citep{guilet_10} even if it develops as a
primary and not as a parasitic instability so that the very absence of
convection for the $27 M_\odot$ progenitor may also be a crucial
factor for violent SASI activity.

Given these conditions, the predominance of the SASI and convection in
model s27.0 and model u8.1, respectively, can be well accounted for,
but the growth conditions can be further connected to the progenitor
structure. The shorter advection time-scale in s27.0 is a direct
result of a significantly higher accretion rate (by a factor of $\sim
\!  4$ at $150 \ \mathrm{ms}$), which is in turn a direct
  consequence of the presence of a dense and rather massive silicon
  shell on top of the iron core (see, e.g., \citealt{woosley_12} for the
  relation between the progenitor structure and the time-dependence of
  $\dot{M}$).  The higher mass accretion rate leads to
a smaller shock stagnation radius (Figure~\ref{fig:shock})
after $30 \ \mathrm{ms}$ post-bounce and thus to a more
narrow post-shock layer as well as higher pre- and post-shock velocities,
which implies a reduction of the accretion time-scale.
Particularly at later stages, the stronger
  gravitational field of the proto-neutron star, whose baryonic mass
  reaches some $1.65 M_\odot$ compared to $1.36 M_\odot$ for u8.1 also
  contributes to the shorter advection time-scale.  According to
Equation~(\ref{eq:chi}), this will reduce the value of $\chi$ and
therefore inhibit the development of convection.

\begin{figure*}
\plottwo{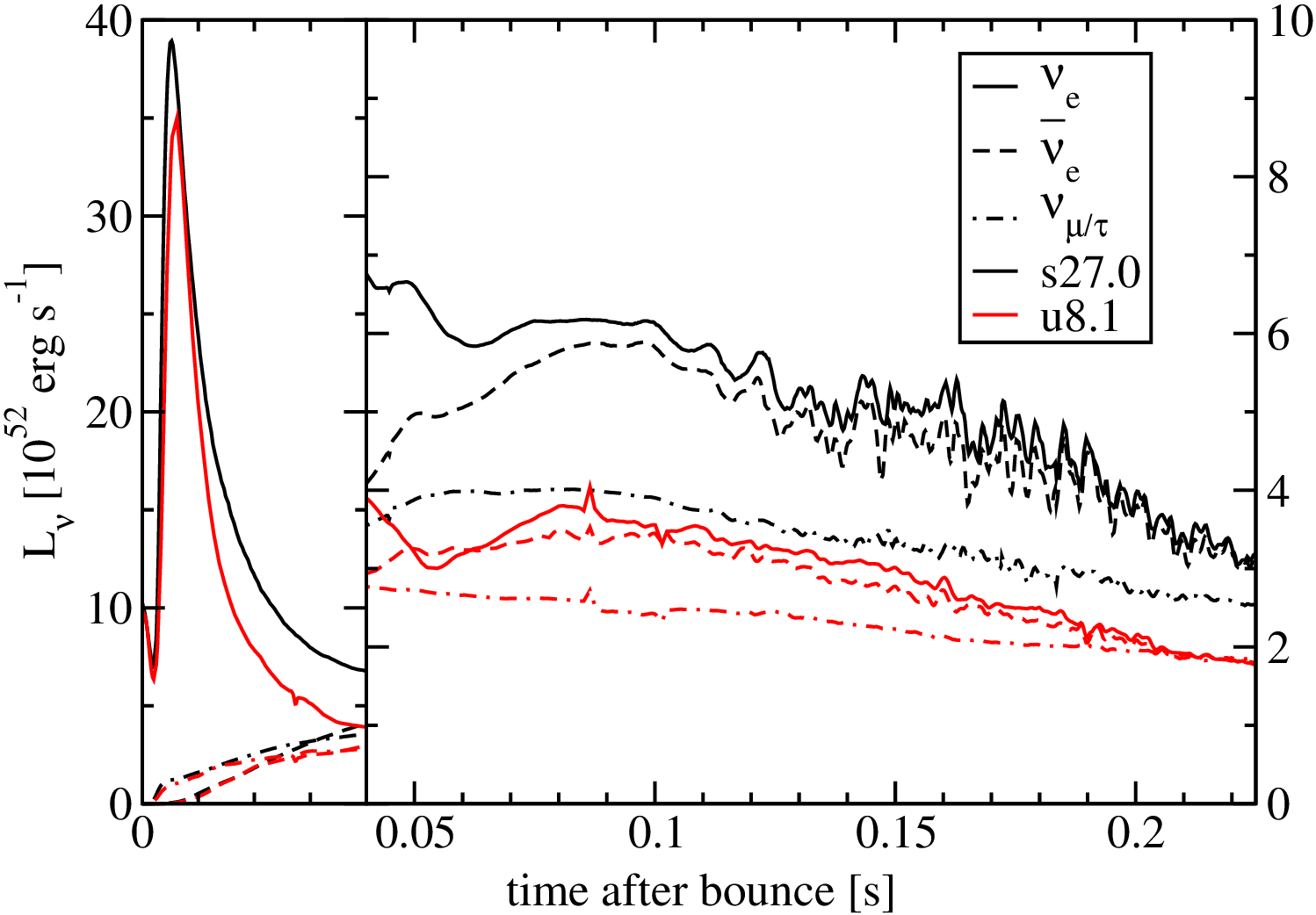}{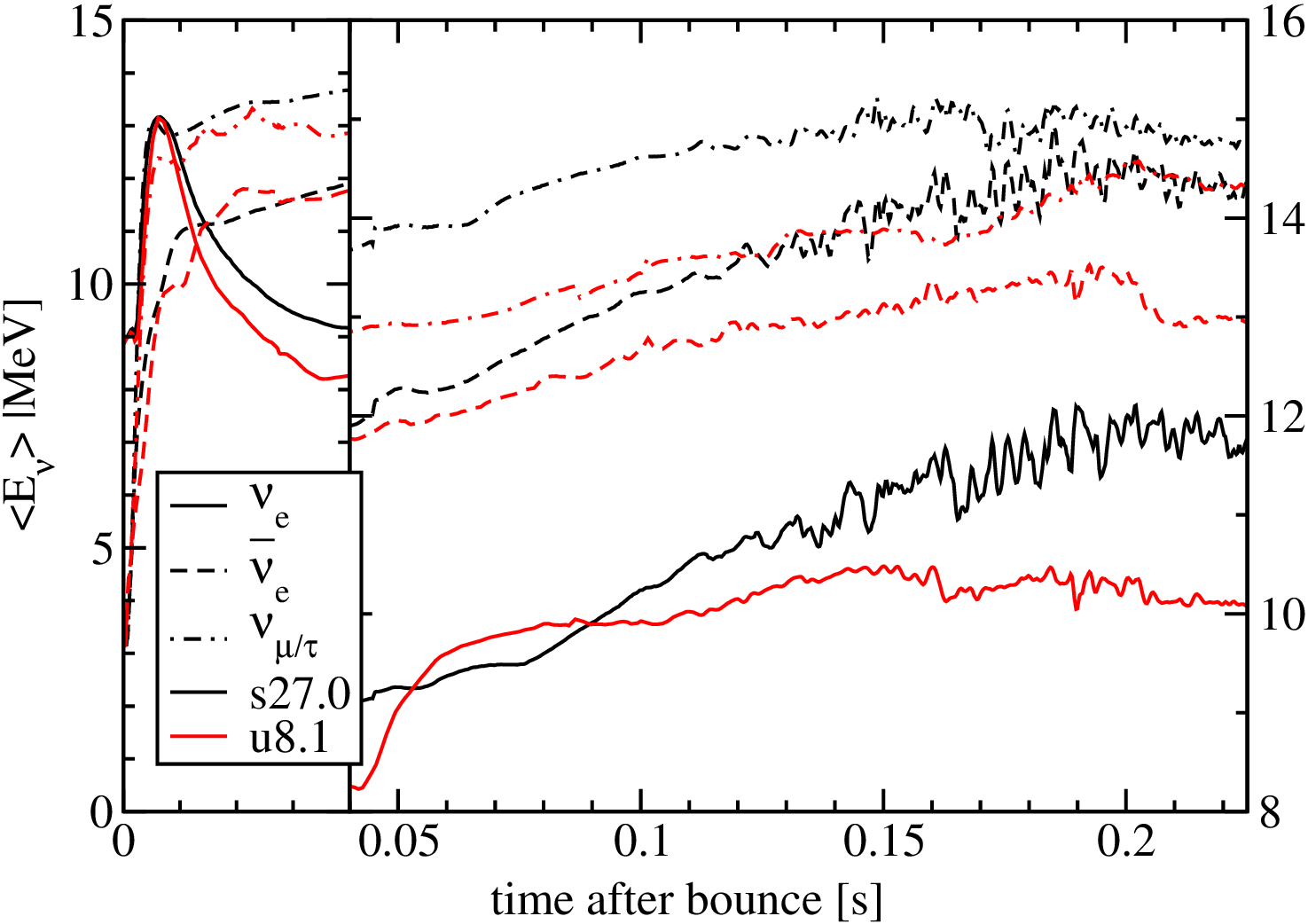}
\caption{Neutrino luminosities (left panel) and mean energies (right
  panel) for s27.0 (black) and u8.1 (red).  Solid, dashed, and
  dash-dotted lines are used for $\nu_e$, $\bar{\nu}_e$, and
  $\nu_{\mu/\tau}$, respectively.  Note that we plot angle-averaged
  quantities extracted at a fiducial observer radius of $400
  \ \mathrm{km}$.
\label{fig:neutrino}
}
\end{figure*}

\begin{figure}
\plotone{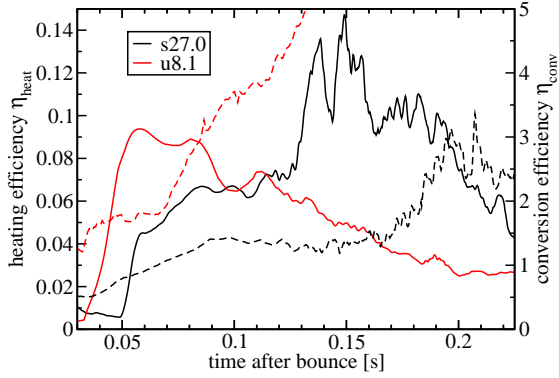}
\caption{Heating efficiency $\eta_\mathrm{heat}$ (solid lines) and
  conversion efficiency $\eta_\mathrm{conv}$ (dashed lines) for s27.0
  (black) and u8.1 (red). All curves have been smoothed using a
  running average over $5 \ \mathrm{ms}$. 
\label{fig:efficencies}}
\end{figure}

Potentially, the larger advection velocities in the gain region could
be compensated by a higher value of $\omega_\mathrm{BV}$ in
Equation~(\ref{eq:chi}), which might result from the stronger neutrino
heating that may be expected because of an enhanced accretion
luminosity (and hence stronger heating) in model
s27.0.   Larger neutrino luminosities and mean
energies are indeed observed for model s27.0 (Figure~\ref{fig:neutrino}),
but what does this imply for the instabilities in the gain region?

To decide this question, we estimate the dependence of the unstable
non-adiabatic gradient, e.g., of the internal energy density
$\epsilon$, on the neutrino emission and the heating efficiency by
means of simple zeroth-order approximation for the solution of the
internal energy equation for an infalling mass shell:
\begin{equation}
\left( \frac{\pd \epsilon}{\pd r} \right)_\mathrm{heating}
\propto \frac{\dot{q}_\nu \tau_\mathrm{adv}}{r_\mathrm{sh}-r_\mathrm{g}}
\propto \frac{\dot{Q}_\nu \tau_\mathrm{adv}}{M_\mathrm{g} \left( r_\mathrm{sh}-r_\mathrm{g} \right)}
\propto \frac{\dot{Q}_\nu }{\dot{M} \left( r_\mathrm{sh}-r_\mathrm{g} \right)}
%\propto \frac{\eta_\mathrm{heat} L}{\dot {M} \left(r_\mathrm{sh}-r_\mathrm{g}\right)}
%\propto L \langle E_\nu^2\rangle \langle v_r \rangle^{-1},
\label{eq:gradient}
\end{equation}
Here, $\dot{q}_\nu$ is the mass-specific neutrino heating rate, which is
approximated in terms of the volume-integrated heating rate $\dot{Q}_\nu$
and the mass in the gain region $M_\mathrm{g} \sim \dot{M}
\tau_\mathrm{adv}$. Equation~(\ref{eq:gradient}) shows that stronger
heating might be balanced by a larger $\dot{M}$, but as $\dot{Q}_\nu$
still depends on $\dot{M}$ both through the post-shock stratification
and the accretion luminosity Equation~(\ref{eq:gradient}) does not
help very much.
  
It is more useful to reformulate Equation~(\ref{eq:gradient}) in terms
of two efficiency parameters $\eta_\mathrm{heat}$ and
$\eta_\mathrm{conv}$ for the neutrino heating and for the conversion
of gravitational energy into electron (anti-)neutrinos.  The familiar
definition of the heating efficiency \citep{marek_09,mueller_12} in
terms of the volume-integrated heating rate $\dot{Q}_\mathrm{heat}$ in
the gain region and the electron neutrino and antineutrino
luminosities\footnote{We use the total angle-integrated neutrino
  energy flux measured at infinity (Figure~\ref{fig:neutrino}) for
  $L_{\nu_e}$ and $L_{\bar{\nu}_e}$. Note that $\eta_\mathrm{heat}$ 
could be interpreted as an effective (non-Rosseland) frequency-averaged optical
depth of the gain region.}  $L_{\nu_e}$ and
$L_{\bar{\nu}_e}$, reads
\begin{equation}
\eta_\mathrm{heat}=\frac{\dot{Q}_\nu}{L_{\nu_e}+L_{\bar{\nu}_e}}.
\end{equation}
As a measure for the efficiency of the conversion of gravitational energy
of the accreted material into neutrinos we define $\eta_\mathrm{conv}$ in terms of
$L_{\nu_e}$, $L_{\bar{\nu}_e}$, and $\dot{M}$ as
\begin{equation}
\eta_\mathrm{conv}=\frac{\left(L_{\nu_e}+L_{\bar{\nu}_e}\right) R}{G M \dot{M}}.
\label{eq:eta_conv}
\end{equation}
Small values of $\eta_\mathrm{conv} <1$ indicate an inefficient
conversion of gravitational energy into accretion luminosity due to
the fast advection of the accreted material through the cooling layer,
whereas $\eta_\mathrm{conv} > 1$ indicates an effective conversion
and an additional enhancement of the electron (anti-)neutrino luminosity
due to the diffusive neutrino flux from the deeper layers of the proto-neutron star.
The unstable gradient manifestly depends on those two efficiencies:
\begin{equation}
\left( \frac{\pd \epsilon}{\pd r} \right)_\mathrm{heating}
\propto \frac{G M \eta_\mathrm{heat} \eta_\mathrm{conv} }{R \left(r_\mathrm{sh}-r_\mathrm{g}\right)}.
\label{eq:gradient2}
\end{equation}

As shown in Figure~\ref{fig:efficencies}, both $\eta_\mathrm{heat}$
and $\eta_\mathrm{conv}$ remain significantly smaller for s27.0 than
for u8.1 before the SASI starts to push the shock outwards at $\sim \!
120 \ \mathrm{ms}$. The smaller heating efficiency
$\eta_\mathrm{heat}$ in s27.0 at early times is a direct consequence
of the rapid advection of material through the gain layer
(Figure~\ref{fig:tadv}), which is compensated by a larger mass in the
gain region at later times.  Particularly large differences are
observed for $\eta_\mathrm{conv}$, which is larger by a factor of $2 \ldots 4$
in model u8.1 prior to the explosion and is therefore by far the most
relevant term in Equation~(\ref{eq:gradient2}). At early times, the
rapid post-shock advection in model s27.0 also results in somewhat
inefficient cooling, whereas neutrino diffusion from the deeper
regions of the proto-neutron star starts to enhance the electron
(anti-)neutrino luminosity rather early for the $8.1 M_\odot$ star and
soon outweighs the accretion luminosity. As a consequence, the
relative difference in luminosity between the two models
(Figure~\ref{fig:neutrino}) is much smaller than the relative difference
  of the accretion rates (Figure~\ref{fig:tadv}). The two terms in
  Equation~(\ref{eq:gradient2}) that could enhance the unstable
  gradient in s27.0 compared to u8.1 are the difference of the shock
  and gain radius in the denominator and the compactness parameter
  $M/R$.  However, the effect of the different value for
  $r_\mathrm{sh}-r _\mathrm{g}$ is more than compensated for by
  different domain of integration in Equation~(\ref{eq:chi}) for the
  parameter $\chi$, and the compactness $M/R$ is only higher by $\sim
  25\%$.

Our analysis thus provides the clue for understanding why
the larger luminosities in model s27.0 are not helpful for fostering
the growth of convective instabilities: The decisive quantity is the
ratio between the luminosity and a fiducial accretion luminosity
$G M \dot{M}/R$ (rather than the luminosity itself), and a
high accretion rate will decrease rather than increase this
``conversion efficiency''.

%% We may thus summarize our analysis of the conditions for SASI-dominated
%% and convectively-dominated flow in model s27.0 and model u8.1 as follows,
%% \begin{itemize}
%% \item The large accretion rate in s27.0 directly decreases
%% $\chi$ because of the larger radial velocities in the post-shock
%% region and the smaller advection time-scale.
%% \item Before shock expansion sets in, the lower heating efficiency
%% in model s27.0 tends to 
%% \end{itemize}

\section{Discussion and Conclusions}
\label{sec:conclusions}
We presented new two-dimensional general relativistic explosion models
for a metal-poor $8.1 M_\odot$ and solar-metallicity $27 M_\odot$
progenitor.  Our successful explosion models for these stars exemplify
two possible regimes for the growth of hydrodynamic instabilities in
core-collapse supernovae.  While the $8.1 M_\odot$ star conforms to
the familiar picture of neutrino-driven convection growing first and
giving rise to low-$\ell$ shock oscillations afterwards as shock
expansion sets in (see, e.g., the $11.2 M_\odot$ and $15 M_\odot$
models of \citealt{marek_09} and \citealt{mueller_12}), the SASI is
the primary instability for the $27 M_\odot$ progenitor.
In this model, we can clearly identify the SASI
and its characteristic features  during the linear growth phase, and find that the flow
remains distinctively different from the convective
$8.1 M_\odot$ model in the non-linear regime.  This ``SASI-dominated''
regime is observed here for the first time in a fully self-consistent
neutrino transport simulation; the purely hydrodynamical models of
\citet{blondin_03}, in which the SASI was initially discovered, and
the SASI models of \citet{scheck_08}, which relied on a fast
contraction of the proto-neutron star, may thus capture the dynamics
in the supernova core better than recently suggested
\citep{burrows_12,murphy_12}.  \emph{Whether convection or the SASI
  emerges as the dominant instability evidently depends on the
  conditions in the accretion flow onto the proto-neutron
  star}. Moreover, the SASI plays a major role during the evolution of
the $27 M_\odot$ progenitor towards an explosion as it already pushes
the shock out to an average radius of $300 \ \mathrm{km}$ before the
runaway condition for neutrino-driven shock expansion is finally
reached and parasitic convective activity becomes very strong. It is
noteworthy that, contrary to the expectations of \citet{fryer_12} and
\citet{belczynsiki_11}, such a SASI-aided explosion can be initiated
at a similarly early stage as in the convectively-dominated model
u8.1, i.e.\ less than $200 \ \mathrm{ms}$ after bounce.

There may be a number of conspiring factors that allow
the $27 M_\odot$ progenitor to reach the SASI-driven regime:
Any effect that leads to higher post-shock advection
  velocities could potentially decide about the character of the
  dominant aspherical instability in the pre-explosion phase. For our
  $27 M_\odot$ model, the following differences to other recent
  two-dimensional first-principle simulations of supernova explosions
  \citep{buras_06b,marek_09,bruenn_09,suwa_10,mueller_12} as well as
  convective models based on a simple light-bulb approximation
  \citep{murphy_08,nordhaus_10,hanke_11,burrows_12,murphy_12}
  contribute -- certainly to varying degrees -- to the short advection
  time-scale:
%% Specifically, the
%% differences between our $27 M_\odot$ model and other recent
%% two-dimensional first-principle simulations of supernova explosions
%% \citep{buras_06b,marek_09,bruenn_09,suwa_10,mueller_12} as well as
%% convective models based on a simple light-bulb approximation
%% \citep{murphy_08,nordhaus_10,hanke_11,burrows_12,murphy_12} may be summarized as
%% follows:
\begin{enumerate}
\item The iron core of the progenitor -- and hence the proto-neutron
  star -- is fairly massive, and is surrounded by a thick and rather
  dense silicon shell. The mass accretion rate therefore remains high
  long after the gain region has formed. This implies a small shock
  stagnation radius and a short advection time-scale, which in turn
  suppresses convective activity (see Equation~\ref{eq:chi}). On the
  other hand, the short advection time-scale is conducive to efficient
  SASI growth (see Equation~\ref{eq:omega}).
\item General relativity further contributes to the short advection
  time-scale as it leads to a more compact neutron star 
  \citep{bruenn_01,mueller_12,lentz_12}, an effect
  that is all the more important as the proto-neutron star is already
  quite massive to begin with. Newtonian models could thus systematically
  miss the SASI-driven regime.
\item The nuclear EoS LS220 produces fairly compact neutron stars in
  agreement with recent radius estimates
  \citep{steiner_10,hebeler_10}, again contributing to fast advection
  through the gain layer (cp.\ \citealt{marek_09} for the dependence
  of the accretion shock radius on the EoS). SASI growth may be inhibited in simulations
  \citep{burrows_06,murphy_08,burrows_12,murphy_12} using the Shen EoS
  \citep{shen_98} with its high value for the symmetry energy
  of $36.9  \ \mathrm{MeV}$.
\item Including the full neutrino transport for all flavors allows the
  proto-neutron star to contract properly, whereas light-bulb models
  and simulations neglecting $\mu$ and $\tau$ neutrinos may suppress
  or underestimate the contraction (and hence the advection
  velocities) considerably. Light-bulb models that disregard neutrino
  diffusion from the neutron star interior neglect this aspect of
  proto-neutron star evolution completely.
\end{enumerate}
With all these elements working in tandem to decrease
the accretion shock radius prior to the onset of multi-dimensional
instabilities, the competition between
buoyancy effects and post-shock advection is heavily tilted towards
the latter, resulting in a suppression of convection
\citep{foglizzo_06}, while the growth of the SASI is accelerated.
The individual contribution of each factor, as well as
  certain balancing effects due to enhanced electron neutrino and
  antineutrino luminosities and spectra, would obviously merit further
  investigation, and cannot be quantified based on a single model.

The realization that the SASI rather than convection can be the
primary instability under certain conditions encountered in
self-consistent 2D supernova simulations opens up several interesting
perspectives: Can a model that is initially dominated by convection
perhaps undergo a late-time transition to the SASI-dominated regime if
the shock retracts sufficiently far?  This could, for example, provide
an explanation for the late explosions of a $15 M_\odot$ star of
\citet{marek_09} and \citet{mueller_12}, which are associated with
strong shock oscillations.  Furthermore, the growth and saturation
behavior of the SASI in 3D become a major issues: Could the parasitic
instabilities that have been proposed as saturation mechanism
\citep{guilet_10} be sufficiently suppressed to allow vigorous SASI
activity in 3D as well?  Could even a partial suppression due to fast
advection explain the development of low-$\ell$ modes, which may be
inhibited under other circumstances because of the forward turbulent
cascade in 3D \citep{hanke_11}? Will the $\ell=1$ sloshing mode be
replaced by a spiral mode that might spin up the neutron star
\citep{blondin_07,fernandez_10} much more effectively than suggested
by recent simulations \citep{wongwathanarat_10,rantsiou_11}?
Will the SASI saturate at lower amplitudes in 3D as
  the kinetic energy is shared among a larger number of modes as
  suggested by \citet{iwakami_09}, or will the kinetic energy
  contained in non-radial motions even become larger as more modes can
  be excited? Will the growth and saturation depend on the initial
  seed perturbations and the rotation rate of the progenitor?  All
these questions will need to be answered \emph{not
    only for restricted parametrized setups}, but for the realistic
conditions actually encountered in self-consistent neutrino
hydrodynamics simulations of a large variety of progenitors.
The future exploration of the SASI and convection will also
  require better methods for discriminating between these two --
  possibly sometimes co-existing -- instabilities in the non-linear
  regime.  The role of SASI and convection in core-collapse
supernovae as well as their mutual interaction are thus bound to
remain fruitful topics in supernova physics.

\acknowledgements AH thanks Candace Joggerst and Stan Woosley for help
with the progenitor star models. This work was supported by the
Deutsche Forschungsgemeinschaft through the Transregional
Collaborative Research Center ``Gravitational Wave Astronomy'' and the
Cluster of Excellence EXC 153 ``Origin and Structure of the Universe''
(http://www.universe-cluster.de).  AH has been supported, in part, by
the DOE Program for Scientific Discovery through Advanced Computing
(SciDAC; DE-FC02-09ER41618), by the US Department of Energy under
grant DE-FG02-87ER40328, by NSF grant AST-1109394, by a Larkin's
Fellowship from Monash University, and by ARC through Future
Fellowship ID FT120100363.  The computations were performed on the IBM
p690 of the Computer Center Garching (RZG), on the Curie supercomputer
of the Grand \'Equipement National de Calcul Intensif (GENCI) under
PRACE grant RA0796, on the Cray XE6 and the NEC SX-8 at the HLRS in
Stuttgart (within project SuperN), on the JUROPA systems at the John
von Neumann Institute for Computing (NIC) in J\"ulich, and on the
Itasca Cluster of the Minnesota Supercomputing Institute.

\bibliography{paper}

\begin{thebibliography}{70}
\expandafter\ifx\csname natexlab\endcsname\relax\def\natexlab#1{#1}\fi

\bibitem[{{Arnett} {et~al.}(1989){Arnett}, {Bahcall}, {Kirshner}, \&
  {Woosley}}]{arnett_89}
{Arnett}, W.~D., {Bahcall}, J.~N., {Kirshner}, R.~P., \& {Woosley}, S.~E. 1989,
  \araa, 27, 629

\bibitem[{{Belczynski} {et~al.}(2011){Belczynski}, {Wiktorowicz}, {Fryer},
  {Holz}, \& {Kalogera}}]{belczynsiki_11}
{Belczynski}, K., {Wiktorowicz}, G., {Fryer}, C., {Holz}, D., \& {Kalogera}, V.
  2011, ArXiv e-prints, 1110.1635

\bibitem[{{Bethe}(1990)}]{bethe_90}
{Bethe}, H.~A. 1990, Rev.~Mod.~Phys., 62, 801

\bibitem[{{Blondin} \& {Mezzacappa}(2006)}]{blondin_06}
{Blondin}, J.~M., \& {Mezzacappa}, A. 2006, \apj, 642, 401

\bibitem[{{Blondin} \& {Mezzacappa}(2007)}]{blondin_07}
------. 2007, \nat, 445, 58

\bibitem[{{Blondin} {et~al.}(2003){Blondin}, {Mezzacappa}, \&
  {DeMarino}}]{blondin_03}
{Blondin}, J.~M., {Mezzacappa}, A., \& {DeMarino}, C. 2003, \apj, 584, 971

\bibitem[{{Bruenn} {et~al.}(2001){Bruenn}, {De Nisco}, \&
  {Mezzacappa}}]{bruenn_01}
{Bruenn}, S.~W., {De Nisco}, K.~R., \& {Mezzacappa}, A. 2001, \apj, 560, 326

\bibitem[{{Bruenn} {et~al.}(2006){Bruenn}, {Dirk}, {Mezzacappa}, {Hayes},
  {Blondin}, {Hix}, \& {Messer}}]{bruenn_06}
{Bruenn}, S.~W., {Dirk}, C.~J., {Mezzacappa}, A., {Hayes}, J.~C., {Blondin},
  J.~M., {Hix}, W.~R., \& {Messer}, O.~E.~B. 2006, {\it J.~Phys.~Conf.~Ser.},
  46, 393

\bibitem[{{Bruenn} {et~al.}(2009){Bruenn}, {Mezzacappa}, {Hix}, {Blondin},
  {Marronetti}, {Messer}, {Dirk}, \& {Yoshida}}]{bruenn_09}
{Bruenn}, S.~W., {Mezzacappa}, A., {Hix}, W.~R., {Blondin}, J.~M.,
  {Marronetti}, P., {Messer}, O.~E.~B., {Dirk}, C.~J., \& {Yoshida}, S. 2009,
  Journal of Physics Conference Series, 180, 012018

\bibitem[{{Buras} {et~al.}(2006{\natexlab{a}}){Buras}, {Janka}, {Rampp}, \&
  {Kifonidis}}]{buras_06b}
{Buras}, R., {Janka}, H.-T., {Rampp}, M., \& {Kifonidis}, K.
  2006{\natexlab{a}}, \aap, 457, 281

\bibitem[{{Buras} {et~al.}(2006{\natexlab{b}}){Buras}, {Rampp}, {Janka}, \&
  {Kifonidis}}]{buras_06_a}
{Buras}, R., {Rampp}, M., {Janka}, H.-T., \& {Kifonidis}, K.
  2006{\natexlab{b}}, \aap, 447, 1049

\bibitem[{{Burrows} {et~al.}(2012){Burrows}, {Dolence}, \&
  {Murphy}}]{burrows_12}
{Burrows}, A., {Dolence}, J.~C., \& {Murphy}, J.~W. 2012, ArXiv e-prints,
  1204.3088

\bibitem[{{Burrows} \& {Fryxell}(1992)}]{burrows_92}
{Burrows}, A., \& {Fryxell}, B.~A. 1992, Science, 258, 430

\bibitem[{{Burrows} \& {Hayes}(1996)}]{burrows_96}
{Burrows}, A., \& {Hayes}, J. 1996, Physical Review Letters, 76, 352

\bibitem[{{Burrows} {et~al.}(1995){Burrows}, {Hayes}, \&
  {Fryxell}}]{burrows_95}
{Burrows}, A., {Hayes}, J., \& {Fryxell}, B.~A. 1995, \apj, 450, 830

\bibitem[{{Burrows} {et~al.}(2006){Burrows}, {Livne}, {Dessart}, {Ott}, \&
  {Murphy}}]{burrows_06}
{Burrows}, A., {Livne}, E., {Dessart}, L., {Ott}, C.~D., \& {Murphy}, J. 2006,
  \apj, 640, 878

\bibitem[{{Chandrasekhar}(1961)}]{chandrasekhar_61}
{Chandrasekhar}, S. 1961, {Hydrodynamic and Hydromagnetic Stability} (Oxford:
  Clarendon)

\bibitem[{{Cordero-Carri{\'o}n} {et~al.}(2009){Cordero-Carri{\'o}n},
  {Cerd{\'a}-Dur{\'a}n}, {Dimmelmeier}, {Jaramillo}, {Novak}, \&
  {Gourgoulhon}}]{cordero_09}
{Cordero-Carri{\'o}n}, I., {Cerd{\'a}-Dur{\'a}n}, P., {Dimmelmeier}, H.,
  {Jaramillo}, J.~L., {Novak}, J., \& {Gourgoulhon}, E. 2009, \prd, 79, 024017

\bibitem[{{Demorest} {et~al.}(2010){Demorest}, {Pennucci}, {Ransom}, {Roberts},
  \& {Hessels}}]{demorest_10}
{Demorest}, P.~B., {Pennucci}, T., {Ransom}, S.~M., {Roberts}, M.~S.~E., \&
  {Hessels}, J.~W.~T. 2010, \nat, 467, 1081

\bibitem[{{Fern{\'a}ndez}(2010)}]{fernandez_10}
{Fern{\'a}ndez}, R. 2010, \apj, 725, 1563

\bibitem[{{Fern{\'a}ndez}(2012)}]{fernandez_12}
------. 2012, \apj, 749, 142

\bibitem[{{Fern{\'a}ndez} \& {Thompson}(2009{\natexlab{a}})}]{fernandez_09b}
{Fern{\'a}ndez}, R., \& {Thompson}, C. 2009{\natexlab{a}}, \apj, 703, 1464

\bibitem[{{Fern{\'a}ndez} \& {Thompson}(2009{\natexlab{b}})}]{fernandez_09a}
------. 2009{\natexlab{b}}, \apj, 697, 1827

\bibitem[{{Foglizzo}(2002)}]{foglizzo_02}
{Foglizzo}, T. 2002, \aap, 392, 353

\bibitem[{{Foglizzo} {et~al.}(2007){Foglizzo}, {Galletti}, {Scheck}, \&
  {Janka}}]{foglizzo_07}
{Foglizzo}, T., {Galletti}, P., {Scheck}, L., \& {Janka}, H.-T. 2007, \apj,
  654, 1006

\bibitem[{{Foglizzo} {et~al.}(2006){Foglizzo}, {Scheck}, \&
  {Janka}}]{foglizzo_06}
{Foglizzo}, T., {Scheck}, L., \& {Janka}, H.-T. 2006, \apj, 652, 1436

\bibitem[{{Fryer} {et~al.}(2012){Fryer}, {Belczynski}, {Wiktorowicz},
  {Dominik}, {Kalogera}, \& {Holz}}]{fryer_12}
{Fryer}, C.~L., {Belczynski}, K., {Wiktorowicz}, G., {Dominik}, M., {Kalogera},
  V., \& {Holz}, D.~E. 2012, \apj, 749, 91

\bibitem[{{Fryer} \& {Warren}(2002)}]{fryer_02}
{Fryer}, C.~L., \& {Warren}, M.~S. 2002, \apjl, 574, L65

\bibitem[{{Gawryszczak} {et~al.}(2010){Gawryszczak}, {Guzman}, {Plewa}, \&
  {Kifonidis}}]{ggpk_10}
{Gawryszczak}, A., {Guzman}, J., {Plewa}, T., \& {Kifonidis}, K. 2010, \aap,
  521, A38

\bibitem[{{Guilet} \& {Foglizzo}(2012)}]{guilet_12}
{Guilet}, J., \& {Foglizzo}, T. 2012, \mnras, 421, 546

\bibitem[{{Guilet} {et~al.}(2010){Guilet}, {Sato}, \& {Foglizzo}}]{guilet_10}
{Guilet}, J., {Sato}, J., \& {Foglizzo}, T. 2010, \apj, 713, 1350

\bibitem[{{Hammer} {et~al.}(2010){Hammer}, {Janka}, \&
  {M{\"u}ller}}]{hammer_10}
{Hammer}, N.~J., {Janka}, H., \& {M{\"u}ller}, E. 2010, \apj, 714, 1371

\bibitem[{{Hanke} {et~al.}(2011){Hanke}, {Marek}, {Mueller}, \&
  {Janka}}]{hanke_11}
{Hanke}, F., {Marek}, A., {Mueller}, B., \& {Janka}, H.-T. 2011, ArXiv
  e-prints, 1108.4355

\bibitem[{{Hebeler} {et~al.}(2010){Hebeler}, {Lattimer}, {Pethick}, \&
  {Schwenk}}]{hebeler_10}
{Hebeler}, K., {Lattimer}, J.~M., {Pethick}, C.~J., \& {Schwenk}, A. 2010,
  Physical Review Letters, 105, 161102

\bibitem[{{Heger} {et~al.}(2012){Heger}, {Woosley}, {Zhang}, \&
  {Joggerst}}]{HWZJ12}
{Heger}, A., {Woosley}, S.~E., {Zhang}, W., \& {Joggerst}, C.~C. 2012, \apj, in
  preparation

\bibitem[{{Herant}(1995)}]{herant_95}
{Herant}, M. 1995, \physrep, 256, 117

\bibitem[{{Herant} {et~al.}(1992){Herant}, {Benz}, \& {Colgate}}]{herant_92}
{Herant}, M., {Benz}, W., \& {Colgate}, S. 1992, \apj, 395, 642

\bibitem[{{Herant} {et~al.}(1994){Herant}, {Benz}, {Hix}, {Fryer}, \&
  {Colgate}}]{herant_94}
{Herant}, M., {Benz}, W., {Hix}, W.~R., {Fryer}, C.~L., \& {Colgate}, S.~A.
  1994, \apj, 435, 339

\bibitem[{{Iwakami} {et~al.}(2008){Iwakami}, {Kotake}, {Ohnishi}, {Yamada}, \&
  {Sawada}}]{iwakami_08}
{Iwakami}, W., {Kotake}, K., {Ohnishi}, N., {Yamada}, S., \& {Sawada}, K. 2008,
  \apj, 678, 1207

\bibitem[{{Iwakami} {et~al.}(2009){Iwakami}, {Kotake}, {Ohnishi}, {Yamada}, \&
  {Sawada}}]{iwakami_09}
------. 2009, \apj, 700, 232

\bibitem[{{Janka}(2001)}]{janka_01}
{Janka}, H.-T. 2001, \aap, 368, 527

\bibitem[{{Janka} \& {M\"uller}(1994)}]{janka_94}
{Janka}, H.-T., \& {M\"uller}, E. 1994, \aap, 290, 496

\bibitem[{{Janka} \& {M\"uller}(1996)}]{janka_96}
------. 1996, \aap, 306, 167

\bibitem[{{Kifonidis} {et~al.}(2003){Kifonidis}, {Plewa}, {Janka}, \&
  {M{\"u}ller}}]{kifonidis_03}
{Kifonidis}, K., {Plewa}, T., {Janka}, H.-T., \& {M{\"u}ller}, E. 2003, \aap,
  408, 621

\bibitem[{{Kifonidis} {et~al.}(2006){Kifonidis}, {Plewa}, {Scheck}, {Janka}, \&
  {M{\"u}ller}}]{kifonidis_06}
{Kifonidis}, K., {Plewa}, T., {Scheck}, L., {Janka}, H.-T., \& {M{\"u}ller}, E.
  2006, \aap, 453, 661

\bibitem[{{Laming}(2007)}]{laming_07}
{Laming}, J.~M. 2007, \apj, 659, 1449

\bibitem[{{Lattimer} \& {Swesty}(1991)}]{lattimer_91}
{Lattimer}, J.~M., \& {Swesty}, F.~D. 1991, {\it Nucl.~Phys.~A}, 535, 331

\bibitem[{{Lentz} {et~al.}(2012){Lentz}, {Mezzacappa}, {Bronson Messer},
  {Liebend{\"o}rfer}, {Hix}, \& {Bruenn}}]{lentz_12}
{Lentz}, E.~J., {Mezzacappa}, A., {Bronson Messer}, O.~E., {Liebend{\"o}rfer},
  M., {Hix}, W.~R., \& {Bruenn}, S.~W. 2012, \apj, 747, 73

\bibitem[{{Marek} \& {Janka}(2009)}]{marek_09}
{Marek}, A., \& {Janka}, H. 2009, \apj, 694, 664

\bibitem[{{M{\"u}ller} {et~al.}(2010){M{\"u}ller}, {Janka}, \&
  {Dimmelmeier}}]{mueller_10}
{M{\"u}ller}, B., {Janka}, H., \& {Dimmelmeier}, H. 2010, \apjs, 189, 104

\bibitem[{{M{\"u}ller} {et~al.}(2012){M{\"u}ller}, {Janka}, \&
  {Marek}}]{mueller_12}
{M{\"u}ller}, B., {Janka}, H.-T., \& {Marek}, A. 2012, ArXiv e-prints,
  1202.0815

\bibitem[{{M\"uller} \& {Janka}(1997)}]{mueller_97}
{M\"uller}, E., \& {Janka}, H.-T. 1997, \aap, 317, 140

\bibitem[{{Murphy} \& {Burrows}(2008)}]{murphy_08}
{Murphy}, J.~W., \& {Burrows}, A. 2008, \apj, 688, 1159

\bibitem[{{Murphy} {et~al.}(2012){Murphy}, {Dolence}, \& {Burrows}}]{murphy_12}
{Murphy}, J.~W., {Dolence}, J.~C., \& {Burrows}, A. 2012, ArXiv e-prints,
  1205.3491

\bibitem[{{Nordhaus} {et~al.}(2012){Nordhaus}, {Brandt}, {Burrows}, \&
  {Almgren}}]{nordhaus_12}
{Nordhaus}, J., {Brandt}, T.~D., {Burrows}, A., \& {Almgren}, A. 2012, \mnras,
  2901

\bibitem[{{Nordhaus} {et~al.}(2010{\natexlab{a}}){Nordhaus}, {Brandt},
  {Burrows}, {Livne}, \& {Ott}}]{nordhaus_10b}
{Nordhaus}, J., {Brandt}, T.~D., {Burrows}, A., {Livne}, E., \& {Ott}, C.~D.
  2010{\natexlab{a}}, \prd, 82, 103016

\bibitem[{{Nordhaus} {et~al.}(2010{\natexlab{b}}){Nordhaus}, {Burrows},
  {Almgren}, \& {Bell}}]{nordhaus_10}
{Nordhaus}, J., {Burrows}, A., {Almgren}, A., \& {Bell}, J. 2010{\natexlab{b}},
  \apj, 720, 694

\bibitem[{{Ohnishi} {et~al.}(2006){Ohnishi}, {Kotake}, \&
  {Yamada}}]{ohnishi_06}
{Ohnishi}, N., {Kotake}, K., \& {Yamada}, S. 2006, \apj, 641, 1018

\bibitem[{{Rampp} \& {Janka}(2002)}]{rampp_02}
{Rampp}, M., \& {Janka}, H.-T. 2002, \aap, 396, 361

\bibitem[{{Rantsiou} {et~al.}(2011){Rantsiou}, {Burrows}, {Nordhaus}, \&
  {Almgren}}]{rantsiou_11}
{Rantsiou}, E., {Burrows}, A., {Nordhaus}, J., \& {Almgren}, A. 2011, \apj,
  732, 57

\bibitem[{{Scheck} {et~al.}(2008){Scheck}, {Janka}, {Foglizzo}, \&
  {Kifonidis}}]{scheck_08}
{Scheck}, L., {Janka}, H.-T., {Foglizzo}, T., \& {Kifonidis}, K. 2008, \aap,
  477, 931

\bibitem[{{Scheck} {et~al.}(2006){Scheck}, {Kifonidis}, {Janka}, \&
  {M{\"u}ller}}]{scheck_06}
{Scheck}, L., {Kifonidis}, K., {Janka}, H.-T., \& {M{\"u}ller}, E. 2006, \aap,
  457, 963

\bibitem[{{Scheck} {et~al.}(2004){Scheck}, {Plewa}, {Janka}, {Kifonidis}, \&
  {M{\"u}ller}}]{scheck_04}
{Scheck}, L., {Plewa}, T., {Janka}, H.-T., {Kifonidis}, K., \& {M{\"u}ller}, E.
  2004, Physical Review Letters, 92, 011103

\bibitem[{{Shen} {et~al.}(1998){Shen}, {Toki}, {Oyamatsu}, \&
  {Sumiyoshi}}]{shen_98}
{Shen}, H., {Toki}, H., {Oyamatsu}, K., \& {Sumiyoshi}, K. 1998, {\it
  Nucl.~Phys.~A}, 637, 435

\bibitem[{{Steiner} {et~al.}(2010){Steiner}, {Lattimer}, \&
  {Brown}}]{steiner_10}
{Steiner}, A.~W., {Lattimer}, J.~M., \& {Brown}, E.~F. 2010, \apj, 722, 33

\bibitem[{{Suwa} {et~al.}(2010){Suwa}, {Kotake}, {Takiwaki}, {Whitehouse},
  {Liebend{\"o}rfer}, \& {Sato}}]{suwa_10}
{Suwa}, Y., {Kotake}, K., {Takiwaki}, T., {Whitehouse}, S.~C.,
  {Liebend{\"o}rfer}, M., \& {Sato}, K. 2010, \pasj, 62, L49+

\bibitem[{{Swesty} {et~al.}(1994){Swesty}, {Lattimer}, \& {Myra}}]{myra_94}
{Swesty}, F.~D., {Lattimer}, J.~M., \& {Myra}, E.~S. 1994, \apj, 425, 195

\bibitem[{{Takiwaki} {et~al.}(2012){Takiwaki}, {Kotake}, \&
  {Suwa}}]{takiwaki_12}
{Takiwaki}, T., {Kotake}, K., \& {Suwa}, Y. 2012, \apj, 749, 98, 1108.3989

\bibitem[{{Thompson} {et~al.}(2003){Thompson}, {Burrows}, \&
  {Pinto}}]{thompson_03}
{Thompson}, T.~A., {Burrows}, A., \& {Pinto}, P.~A. 2003, \apj, 592, 434

\bibitem[{{Thompson} {et~al.}(2005){Thompson}, {Quataert}, \&
  {Burrows}}]{thompson_05}
{Thompson}, T.~A., {Quataert}, E., \& {Burrows}, A. 2005, \apj, 620, 861

\bibitem[{{Torrence} \& {Compo}(1998)}]{torrence_98}
{Torrence}, C., \& {Compo}, G.~P. 1998, Bulletin of the American Meteorological
  Society, 79, 61

\bibitem[{{Wongwathanarat} {et~al.}(2010){Wongwathanarat}, {Janka}, \&
  {M{\"u}ller}}]{wongwathanarat_10}
{Wongwathanarat}, A., {Janka}, H., \& {M{\"u}ller}, E. 2010, \apjl, 725, L106

\bibitem[{{Woosley} \& {Heger}(2012)}]{woosley_12}
{Woosley}, S.~E., \& {Heger}, A. 2012, \apj, 752, 32

\bibitem[{{Woosley} {et~al.}(2002){Woosley}, {Heger}, \& {Weaver}}]{woosley_02}
{Woosley}, S.~E., {Heger}, A., \& {Weaver}, T.~A. 2002, {\it Rev.~Mod.~Phys.},
  74, 1015

\bibitem[{{Woosley} \& {Weaver}(1995)}]{woosley_95}
{Woosley}, S.~E., \& {Weaver}, T.~A. 1995, \apjs, 101, 181

\bibitem[{{Yamasaki} \& {Yamada}(2007)}]{yamasaki_07}
{Yamasaki}, T., \& {Yamada}, S. 2007, \apj, 656, 1019

\end{thebibliography}

\end{document}